\documentclass[pre,superscriptaddress, floatfix]{revtex4-1}
\usepackage{amsmath,amssymb}
\usepackage{dcolumn}
\usepackage{graphicx,color,hyperref}
\usepackage{wrapfig}
\usepackage{bm}
\usepackage{enumerate}
\usepackage{amssymb, amsthm, amsmath}

\usepackage{url}
\usepackage{ulem}
\newcommand{\rd}[1]{{\textcolor{red}{ #1}}}


\begin{document}
\title{Integrative analysis of ATAC-seq and RNA-seq for cells \texorpdfstring{\\infected by human T-cell leukemia virus type 1}{} }

\author{Azusa Tanaka}
\affiliation{Department of Human Genetics, Graduate School of Medicine, The University of Tokyo}

\author{Yasuhiro Ishitsuka}
\affiliation{Institute of Mathematics for Industry, Kyushu University}

\author{Hiroki Ohta}
\affiliation{Department of Human Sciences, Obihiro University of Agriculture and Veterinary Medicine}

\author{\\Norihiro Takenouchi}
\affiliation{Department of Microbiology, Kansai Medical University}

\author{Masanori Nakagawa}
\affiliation{Kyoto Prefectural University of Medicine}

\author{Ki-Ryang Koh}
\affiliation{Department of Hematology, Osaka General Hospital of West Japan Railway Company}

\author{Chiho Onishi}
\affiliation{Laboratory of Ultrastructural Virology, Institute for Life and Medical Sciences, Kyoto University}

\author{\\Hiromitsu Tanaka}
\affiliation{Department of Biophysics, Graduate school of Science, Kyoto University}
\affiliation{Department of Developmental Biology, Graduate School of Medicine, Chiba University}

\author{Akihiro Fujimoto}
\affiliation{Department of Human Genetics, Graduate School of Medicine, The University of Tokyo}

\author{Jun-ichirou Yasunaga}
\affiliation{Department of Hematology, Rheumatology and Infectious Disease, Faculty of Life Sciences, Kumamoto University}

\author{Masao Matsuoka}
\affiliation{Department of Hematology, Rheumatology and Infectious Disease, Faculty of Life Sciences, Kumamoto University}

\date{\today}

\begin{abstract}
  Human T-cell leukemia virus type 1 (HTLV-1) causes adult T-cell leukemia (ATL) and HTLV-1-associated myelopathy (HAM) after a long latent period in a fraction of infected individuals. These HTLV-1-infected cells typically have phenotypes similar to that of CD4${^+}$ T cells, but the cell status is not well understood.
  To extract the inherent information of HTLV-1-infected CD4$^+$ cells, we integratively analyzed the ATAC-seq and RNA-seq data of infected cells. Compared to CD4${^+}$ T cells from healthy donors, we found anomalous chromatin accessibility in HTLV-1-infected CD4${^+}$ cells derived from ATL cases in terms of location and sample-to-sample fluctuations in open chromatin regions.
Further, by focusing on systematically selected genes near the open chromatin regions,  all the gene expressions in ATL cases were found to be distinct from those of healthy CD4$^+$ T cells.
Based on a further analysis of chromatin accessibility, we detected TLL1 (Tolloid Like 1) as one of the key genes that exhibit unique gene expressions in ATL cases. A luciferase assay indicated that TLL1 has a strong regulatory effect on TGF-$\beta$. 
Overall, this study provides  results about the status of HTLV-1 infected cells, which are qualitatively consistent across the different scales of chromatin accessibility, transcription, and  immunophenotype.

 \end{abstract}

\maketitle

\section{Introduction}
It has been statistically estimated that there are more than $300,000$ types of mammalian host viruses \cite{doi:10.1128/mBio.00598-13}.
Among the many viruses that have been discovered, only a few have been reported to cause cancers, such as the DNA virus human papillomavirus (HPV) and the RNA virus hepatitis C virus (HCV) \cite{KrumpYou2018}. One, human T-cell leukemia virus (HTLV-1), is a retrovirus, oncogenic virus, and
estimated to infect approximately $10$ million people worldwide \citep{Gessain2012}.

Adult T-cell leukemia (ATL) and HTLV-1-associated myelopathy (HAM) are both associated with prior infection with HTLV-1. However, these two diseases have different clinical and pathological presentations \citep{Matsuoka2007, Bangham2023}.
The genes encoded by HTLV-1, such as HBZ (HTLV-1 basic leucine zipper factor) and \textit{Tax},
have been reported to affect important signaling pathways involved in cell proliferation, apoptosis, and infectivity \citep{Yamagishi2018}.
In particular, HBZ is maintained in all ATL cases and functions as both a protein and RNA \citep{Satou2006,Tanaka-Nakanishi2014,Mitobe2015, Ma2016}.
Recent studies have elucidated that in ATL cells, genomic mutations are highly enriched in T cell-related pathways, such as NF-${\kappa}$B, and typically activate the pathways \citep{Kataoka2015,Kataoka2016_2}.
Furthermore, it has been frequently observed in ATL cases that the aberrant expression of programmed cell death 1-ligand 1 (PD-L1) is caused by disruption of the PD-L1 3'-untranslated region (UTR) \citep{Kataoka2016_1}.

Several questions about these diseases at the genomic scale remain, including how the chromatin structure of ATL cells differs from that of CD4$^+$ T cells derived from healthy donors and how this difference influences transcription and translation to finally cause symptoms.
In general, cellular phenotypes are largely affected by gene expressions that are strongly correlated with the epigenetic mechanisms occurring in chromatin. 
To understand the epigenetic mechanisms, it is important to understand how human DNA is packed and chemically modified in the nucleus, which can be quantified by measuring chromatin accessibility.

In this paper, we study the relationship between chromatin accessibility and transcription in HTLV-1-infected cells at the whole genome level using Assay for Transposase-Accessible Chromatin using sequencing (ATAC-seq) \citep{ATAC1} and RNA sequencing (RNA-seq) data.
We performed a comparative analysis of HTLV-1 infected CD4$^+$ cells from ATL cases, HAM cases, and CD4${^+}$ T cells from healthy donors (healthy CD4${^+}$ T cells) based mainly on our previously developed algorithm of systematic clustering \citep{Tanaka2020}. 

Our analysis shows that CD4${^+}$ cells derived from ATL cases have anomalous properties in terms of the locations and sample-to-sample fluctuations of open chromatin regions compared with healthy CD4${^+}$ T cells.
Additionally, genes selected by our systematic clustering algorithm based on the immunophenotype had distinct expressions between ATL cells and healthy CD4${^+}$ T cells.
Using our systematic clustering algorithm, we also found a relationship between chromatin accessibility and immunophenotype that suggests some ATL cases approach several types of myeloid cells.

Finally, we detected TLL1 (Tolloid Like 1) as one of a few genes having anomalous expressions in ATL cases. A luciferase assay found TLL1 isoforms, depending on the types of its isoforms, regulate differently the maturation of TGF-$\beta$ (transforming growth factor $\beta$), which is known to play important roles in cancer progression.

\section{Results}

\begin{figure}[t!]

   \centering
    
  \includegraphics[trim = {0 20pt 0 30pt}, clip, width = 0.8\linewidth]{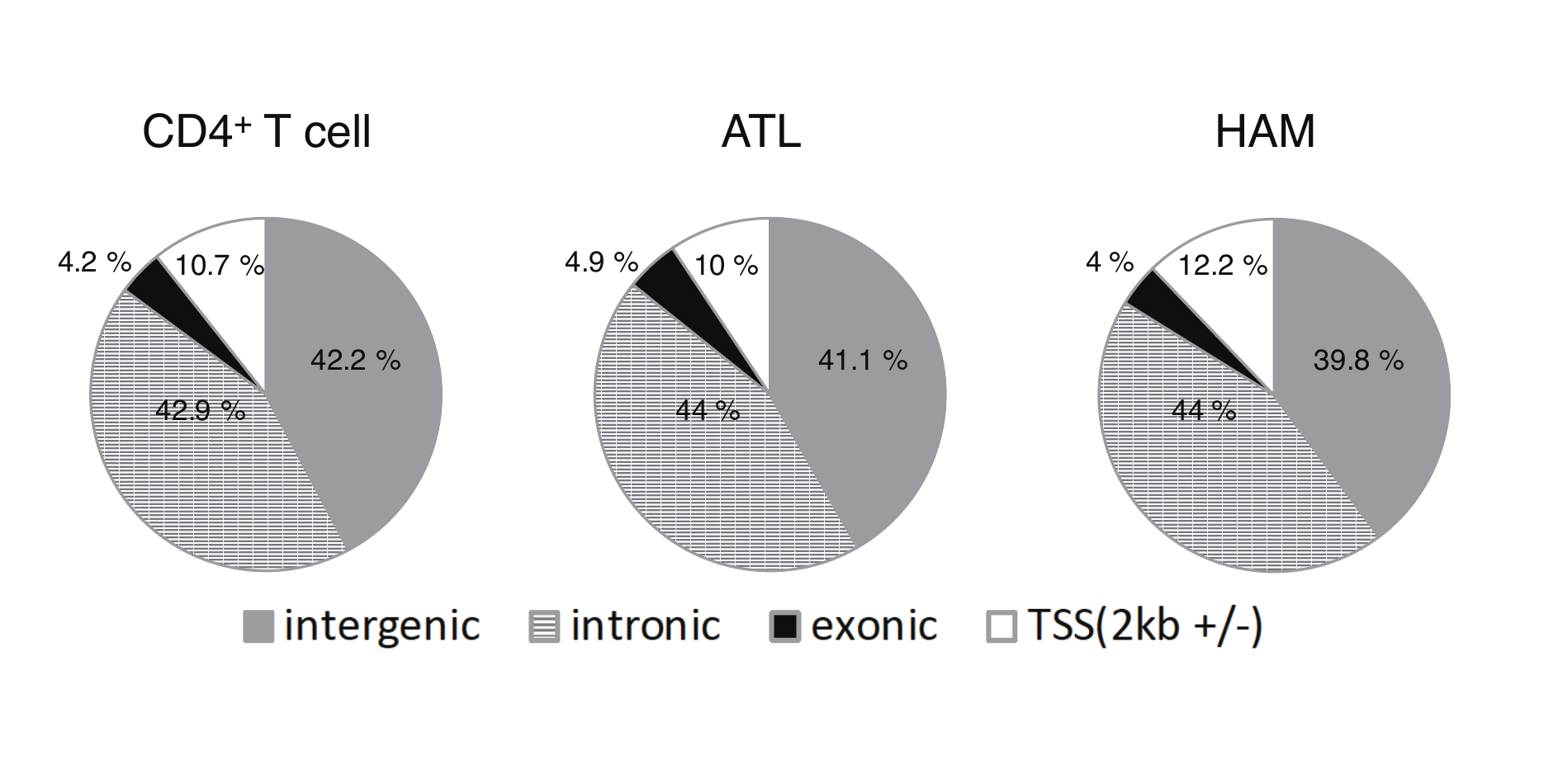}
\caption{The genetic (intergenic/intronic/exonic) annotation of ATAC-seq peaks, which quantify open chromatin regions. TSS(2kb $+/-$) corresponds to $- 2000$ to $2000$ base pairs from a transcription start site.}
\label{fig:ATAC_peaks}
\end{figure}

\subsection{Chromatin accessibility: whole view of the genome} \label{Rsection1}


\begin{figure}[t!]

   \begin{flushleft}
    (a) 
   \end{flushleft}
   \centering\label{fig:ATAC_peak_annotation}\includegraphics[trim = 0 50 0 0, clip, keepaspectratio, width=0.8\linewidth]{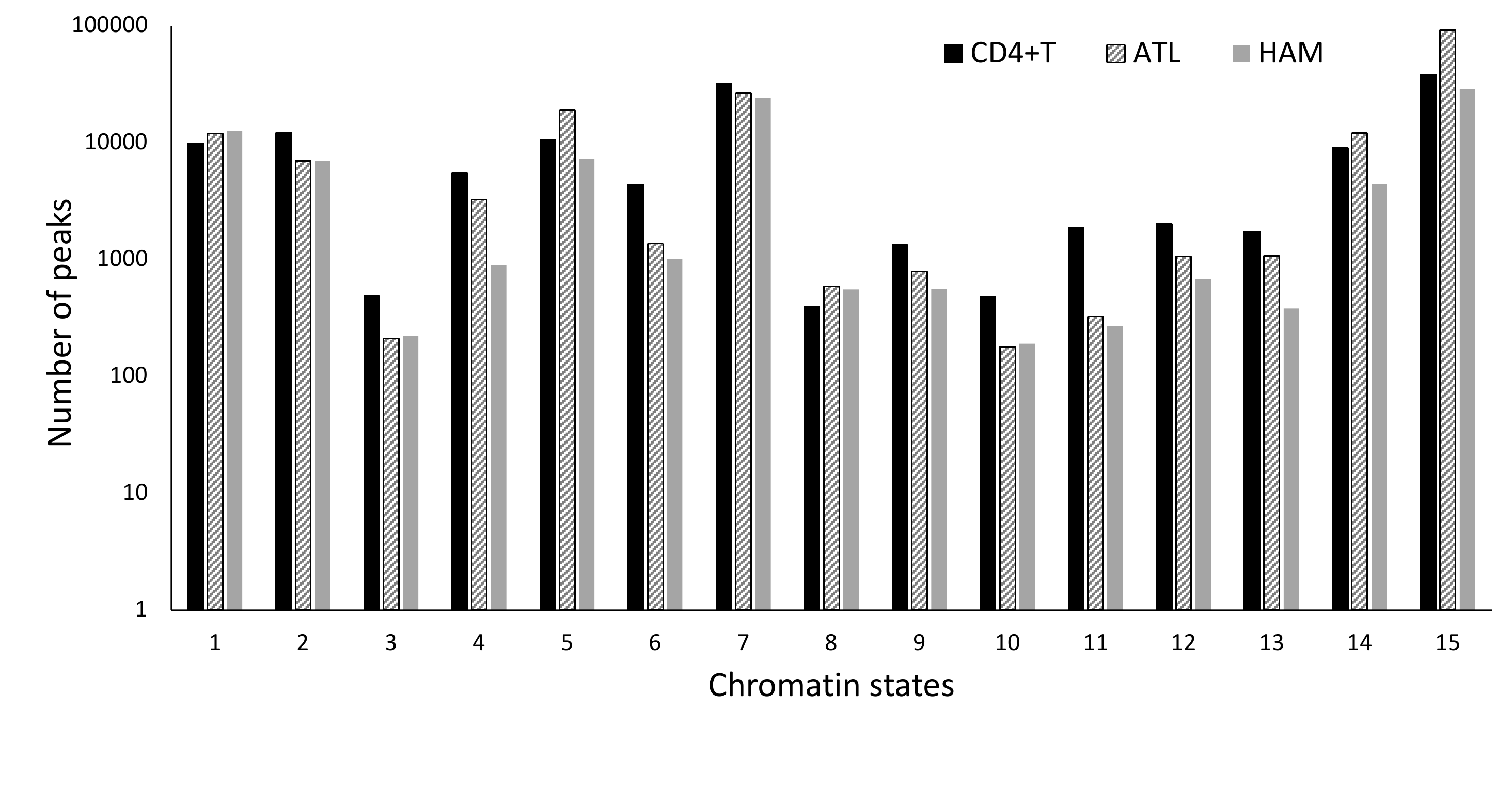}

\begin{flushleft}
    {(b)}
\end{flushleft}
\begin{tabular}{|c|c|c|c|c|c|c|}
\hline
State No. & Description & $N_{\mathrm{CD4^+T}}$  & $N_{\mathrm{ATL}}$ & $N_{\mathrm{HAM}}$ & \rule[0pt]{0pt}{3.5ex}$\dfrac{N_{\mathrm{ATL}}}{N_{\mathrm{CD4^+T}}}$ &$\dfrac{N_{\mathrm{HAM}}}{N_{\mathrm{CD4^+T}}}$ \\[1em]
\hline   
1 & Active TSS & 9939 & 12084 & 12691 & 1.22 & 1.28\\
2 & Flanking Active TSS & 12202	& 7015 & 6983 & 0.575 & 0.572\\
3 & Transcr. at gene 5' and 3' & 489 & 213 & 224 & 0.436 & 0.458\\
4 & Strong transcription & 5525 & 3278 & 893 & 0.593 & 0.162\\
5 & Weak transcription & 10659 & 18946 & 7269 & 1.78 & 0.682\\
6 & Genic enhancers & 4431 & 1374 & 1017 & 0.310 & 0.230\\
7 & Enhancers & 32260 & 26597 & 24283 & 0.824 & 0.753\\
8 & ZNF genes \& repeats & 401 & 595 & 559 & 1.48 & 1.39\\
9 & Heterochromatin & 1340 & 794 & 565 & 0.593 & 0.422\\
10 & Bivalent/Poised TSS & 478 & 181 & 192 & 0.379 & 0.402\\
11 & Flanking Bivalent TSS/Enh & 1894 & 326 & 268 & 0.172 & 0.141\\
12 & Bivalent Enhancer & 2031 & 1074 & 682 & 0.529 & 0.336\\
13 & Repressed PolyComb & 1746 & 1077 & 382 & 0.617 & 0.219\\
14 & Weak Repressed PolyComb & 9086 & 12178 & 4445 & 1.34 & 0.489\\
15 & Quiescent/Low & 38566 & 92307 & 28746 & 2.39 & 0.745\\
\hline
\end{tabular}
\caption{(a) The number of ATAC-seq peaks (vertical axis) vs. indices (horizontal axis) classified by each functional annotation from $1$ to $15$,  as shown in (b). (b) The number of ATAC-seq peaks classified into each functional annotation for healthy CD4$^+$T cells, ATL cells, and HAM cells. $N_{\mathrm{CD4^+T}}$, $N_{\mathrm{ATL}}$, and $N_{\mathrm{HAM}}$: peak number of healthy CD4$^+$T, ATL, and HAM, respectively. 
Note that each peak quantifies an open chromatin region.}
\label{fig:ATAC_annot_ratio}
\end{figure}

The landscape of chromatin accessibility provides useful information for understanding the mechanisms that govern cell-type-specific gene expressions.
Preliminarily, we overview the chromatin accessibility characterized by the ATAC-seq of healthy CD4$^+$ T cells, ATL cells, and HAM cells.

To obtain the chromatin accessibility landscape, we performed ATAC-seq on HTLV-1-infected CD4$^+$ cells obtained from the peripheral blood of $29$ ATL and $6$ HAM cases.
All samples selected for the ATAC-seq library preparation were at least $98$\% HTLV-1-infected cells. The ATAC-seq libraries were sequenced with an average of $44$ million reads, resulting in a dataset comprising of $1.3$ billion and $556$ million sequenced reads for ATL and HAM, respectively. The data quality was high in all cases, with mitochondrial read rates of $7.5$\% for ATL and $7.3$\% for HAM.
For a comparison, we used ATAC-seq datasets of CD4${^+}$ T cell samples from $5$ healthy donors and downloaded from GEO accession GSE74912 \citep{Howard}. 

To identify genome-wide accessible chromatin regions, for each of the three groups, we concatenated ATAC-seq reads for the different samples, where the sample number was $29$ for ATL, $6$ for HAM, and $5$ for healthy CD4${^+}$ T cells.
As explained in the Materials and Methods, we randomly selected $100$M reads from the concatenated data of each group. 
We used the MACS2 algorithm to select the locations of peaks to quantify the open chromatin regions from the ATAC-seq datasets \citep{Zhang2008}, finding a total of $178811$, $89972$, and $131609$ peaks in ATL, HAM, and healthy CD4${^+}$ T cells, respectively. 

The ENCODE consortium shows that $10$\% of peaks are localized in near transcription start sites (TSSs), whereas the remaining $90$\% of peaks are mapped
nearly equally to intronic and intergenic regions \citep{Thurman2012}. Consistent with these data, as shown in Fig.\ \ref{fig:ATAC_peaks}, about $10$\% of the ATAC-seq peaks are overlapped with the TSSs and their surrounding regions, whereas the majority of ATAC-seq peaks (about $85$\%) of healthy CD4${^+}$ T cells, ATL cells, and HAM cells reside in intergenic or intronic regions.

To determine the functional roles of the peaks in HTLV-1-infected cells and healthy CD4${^+}$ T cells, we computed the overlapping ratio of these regions with specific genomic features, such as active TSS, enhancers, heterochromatin, etc.
To assign a genomic feature to all genomic positions, we assumed a chromHMM $15$-state model obtained from (\url{https://egg2.wustl.edu/roadmap/web_portal/chr_state_learning.html}). We used the data of E043 for healthy CD4${^+}$ T cells and E037 for HTLV-1-infected CD4$^+$ cells. Note that E037 is the  model of CD4$^+$ memory T cells because a majority of ATL cases has been reported to show CD45RO$^+$, which is consistent with CD4$^+$ memory T cells \citep{Richardson1990}.

As shown in Fig.\ \ref{fig:ATAC_annot_ratio}, for HTLV-1-infected CD4$^+$ cells, the number of peaks compared with healthy CD4${^+}$ T cells was proportionally less in categories related to Enhancer $(6, 7, 11, 12)$ and Heterochromatin $(9)$, but it was higher for the category of Quiescent/Low $(15)$ only for ATL cases.

This observation suggests that compared with healthy CD4$^+$ T cells, distinct enhancer mechanisms in HTLV1-infected cases are correlated with the distinct chromatin structures. 


\begin{figure}[t]
\begin{center}
\begin{flushleft}
     {(a)} 
\end{flushleft}
\begin{minipage}[b]{\linewidth}
     \centering

\includegraphics[keepaspectratio,clip,
trim = 0 20 0 40, width=0.5\linewidth]{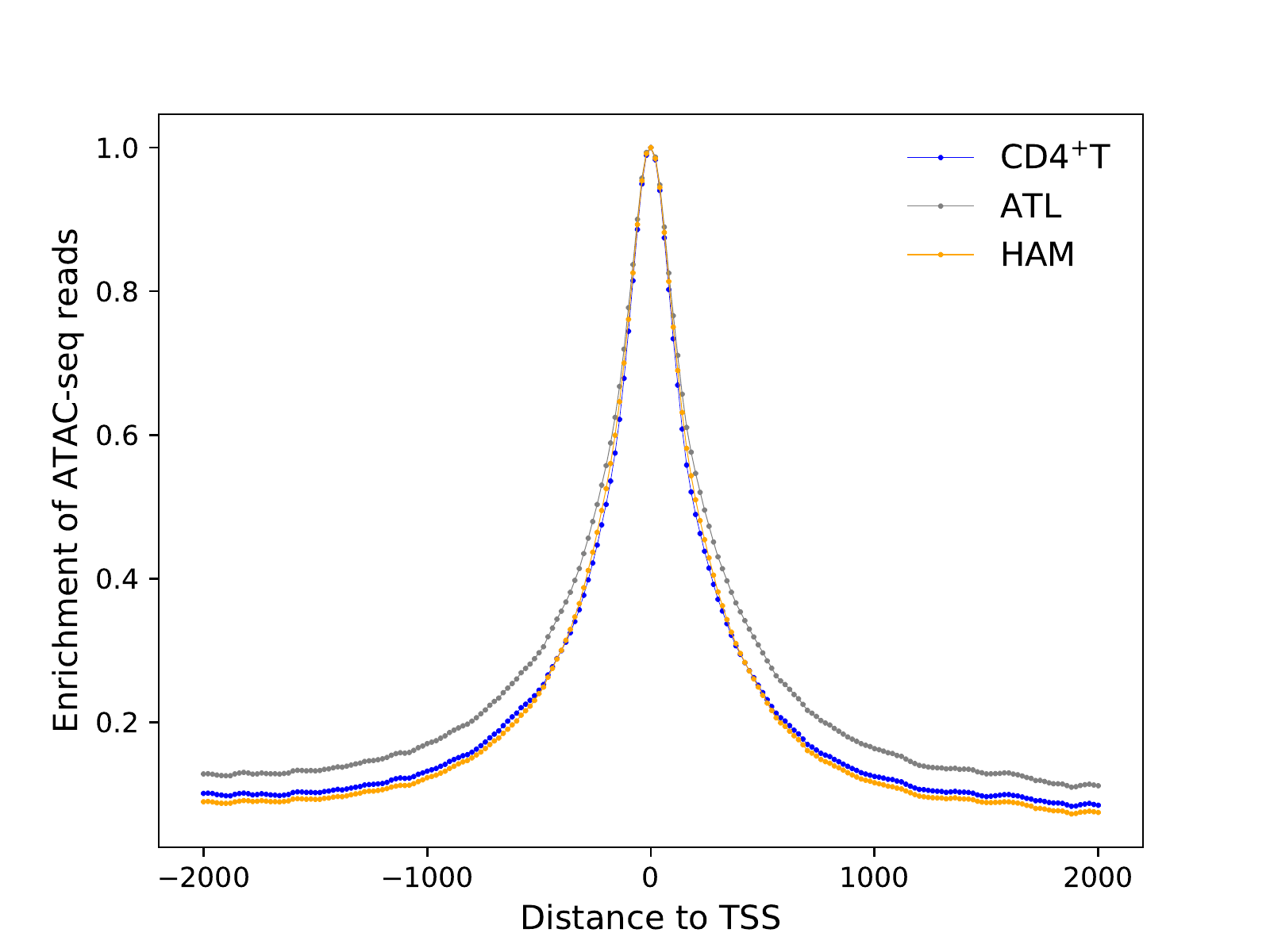}
  \end{minipage}

\vspace{15pt}
\begin{flushleft}
    {(b)}
\end{flushleft}
  \begin{minipage}[b]{\linewidth}
     \centering

    \includegraphics[
    trim = 20 20 50 30, clip,
    keepaspectratio, width = 0.32\linewidth]{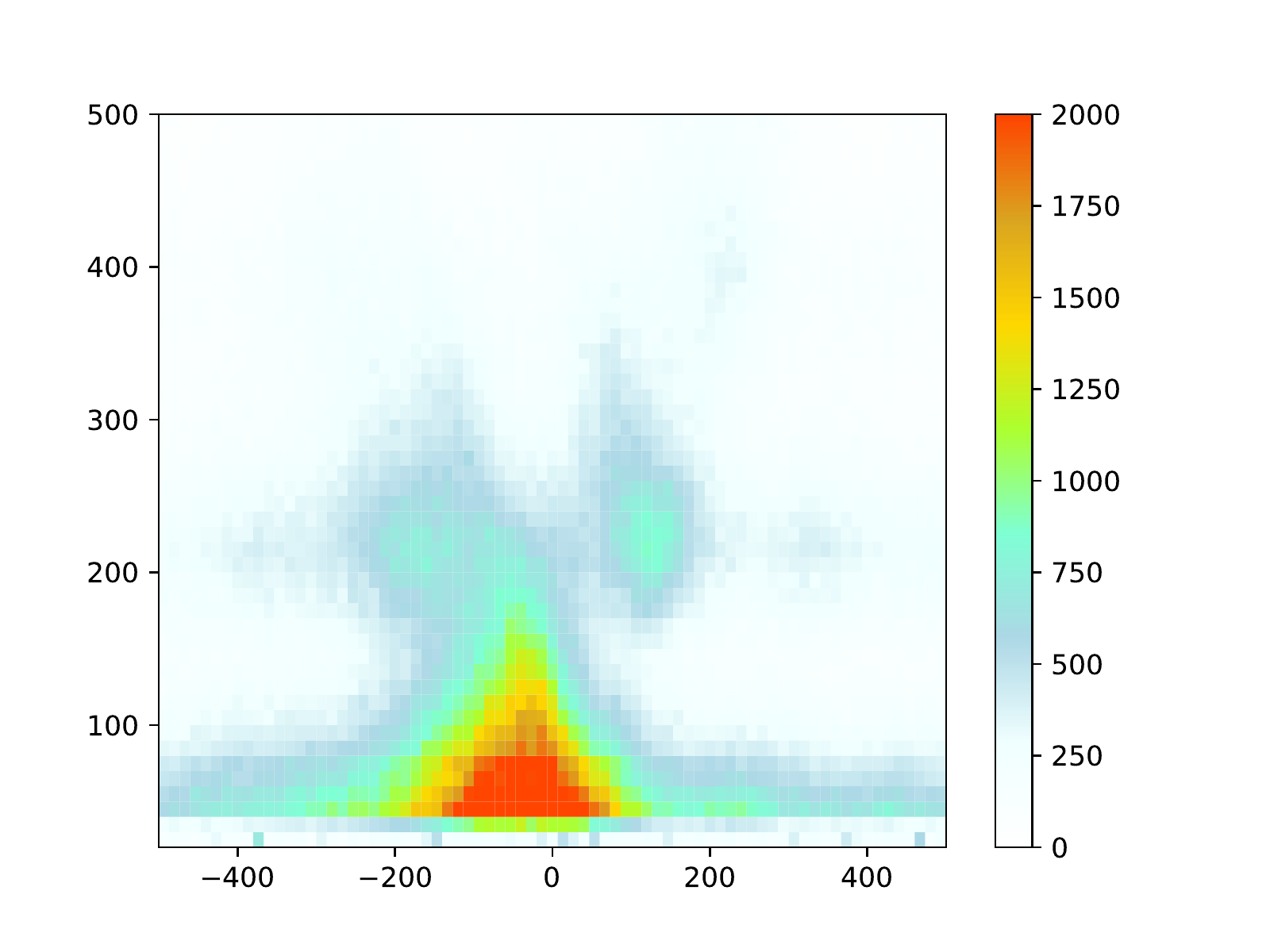}   
    \includegraphics[
    trim = 20 20 50 30, clip,
    keepaspectratio, width = 0.32\textwidth]{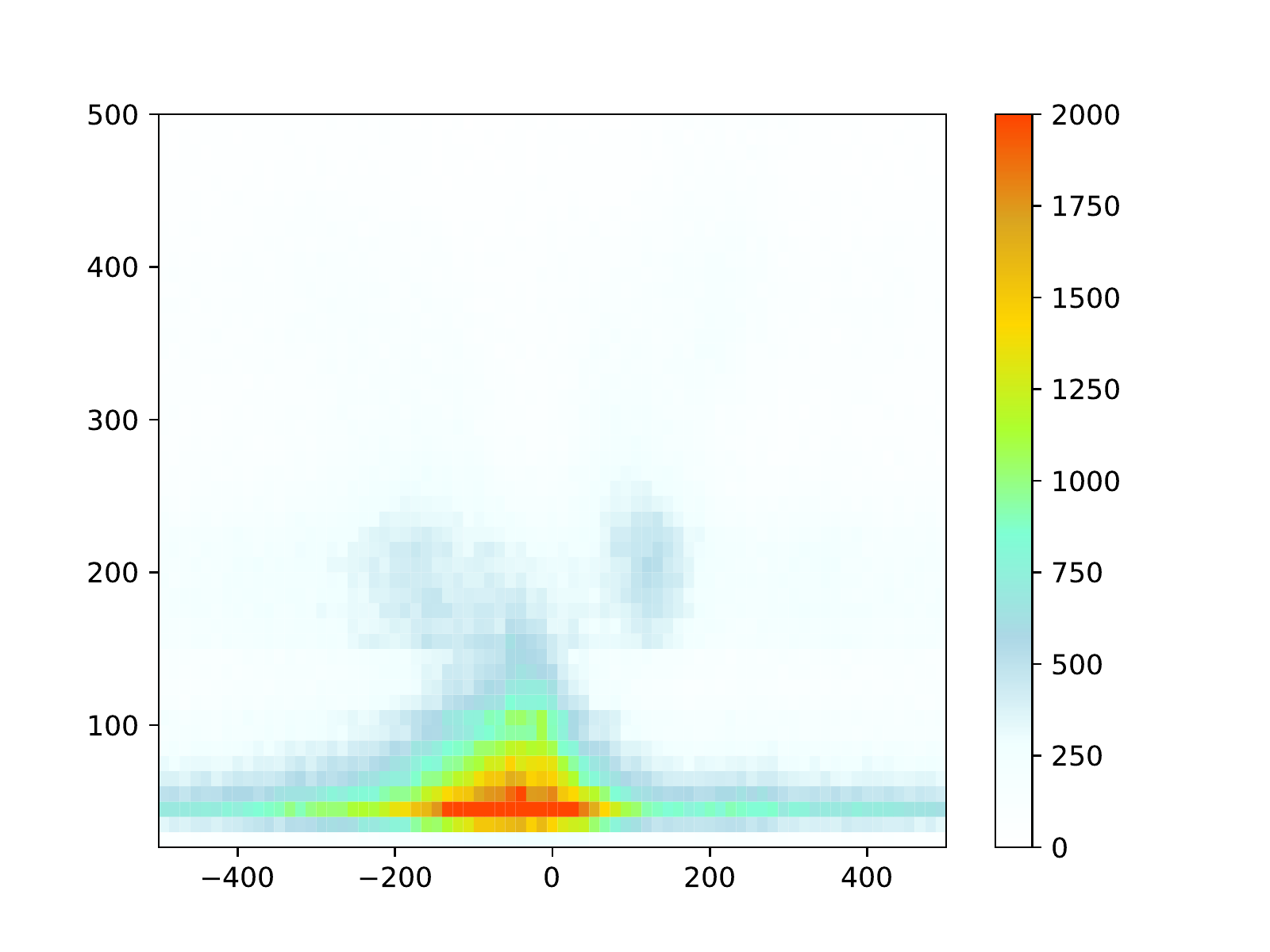}   
    \includegraphics[
    trim = 20 20 50 30, clip,
    keepaspectratio, width = 0.32\textwidth]{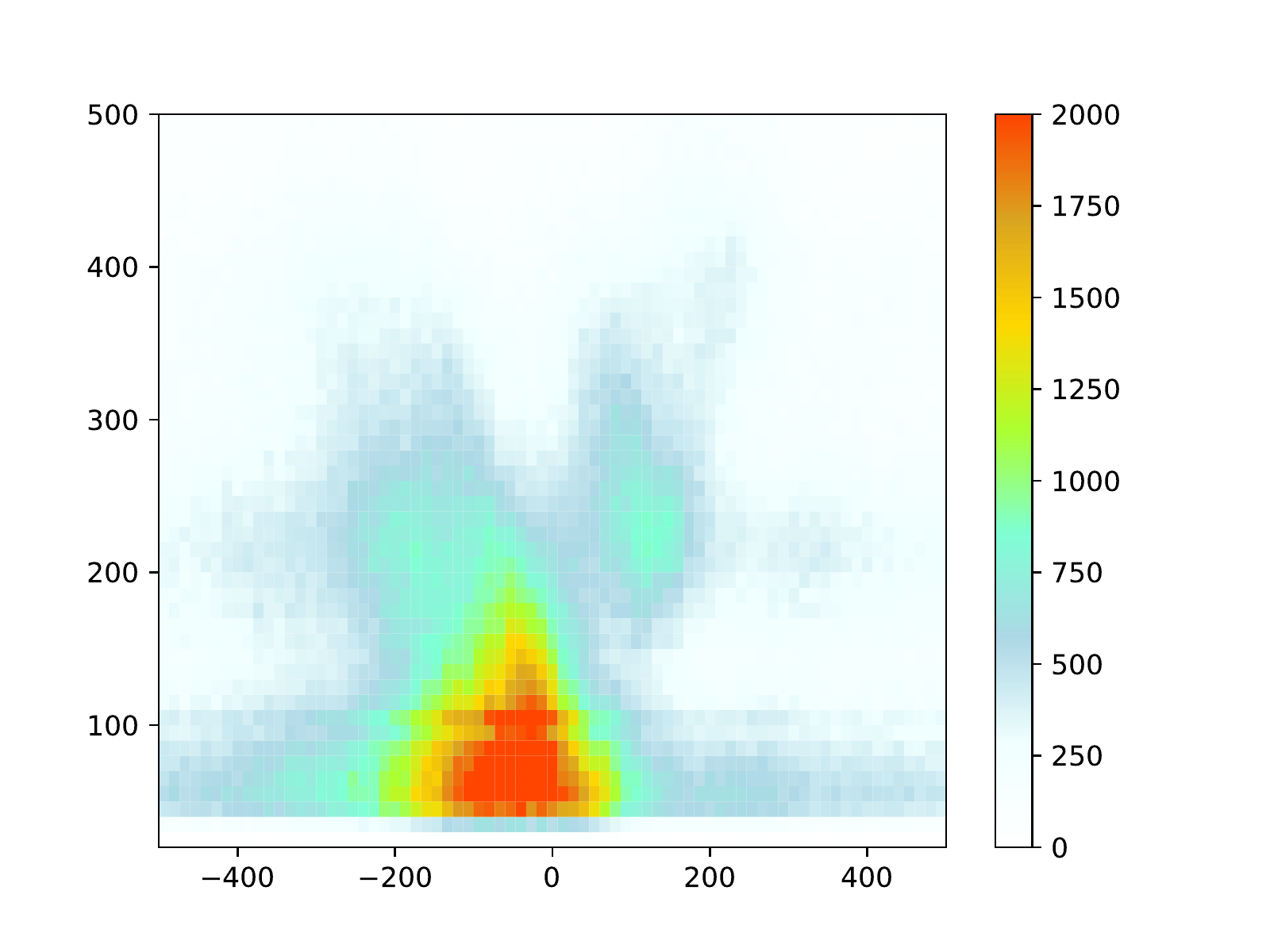}
  \end{minipage}

\end{center}


\caption{
(a) The normalized number $\widetilde{\rho}_{\nu}(z)$ of reads at position $z$ from the TSS for type $\nu$ as defined in \eqref{eq:1d_TSS_quantity}.
     Note that position $z$ takes the value $\{ -2000+20k \mid 0\le k \le 200\}$, where $k$ is an integer. 
(b) 
A histogram (heat map) $F^{\Delta,\xi}_{\nu} (z,\ell)$ of the fragment length $\ell$ of the ATAC-seq data at position $z$ from the TSSs as defined in \eqref{eq:2d_TSS_quantity}. (left) CD4$^+$T, (center) ATL, and (right) HAM. The graphs are plotted for $z \in \{n\Delta \mid -37 \le n \le 37\}$ and $\ell \in \{n \xi \mid 0 \le n \le 49 \}$, where 
    $\Delta = 1000/75$ and $\xi = 10$.}
 
\label{fig:around_TSS}
\end{figure}

\subsection{Increased chromatin accessibility around transcriptional start sites (TSSs) in ATL} \label{Rsection3}

\begin{figure}[t]
\begin{center}
 
  \begin{minipage}[b]{0.45\linewidth}
     \centering
    \begin{flushleft}
        {(a) All regions}
    \end{flushleft}

    \includegraphics[ 
    trim = 0 0 0 40, clip,
    width=\linewidth]{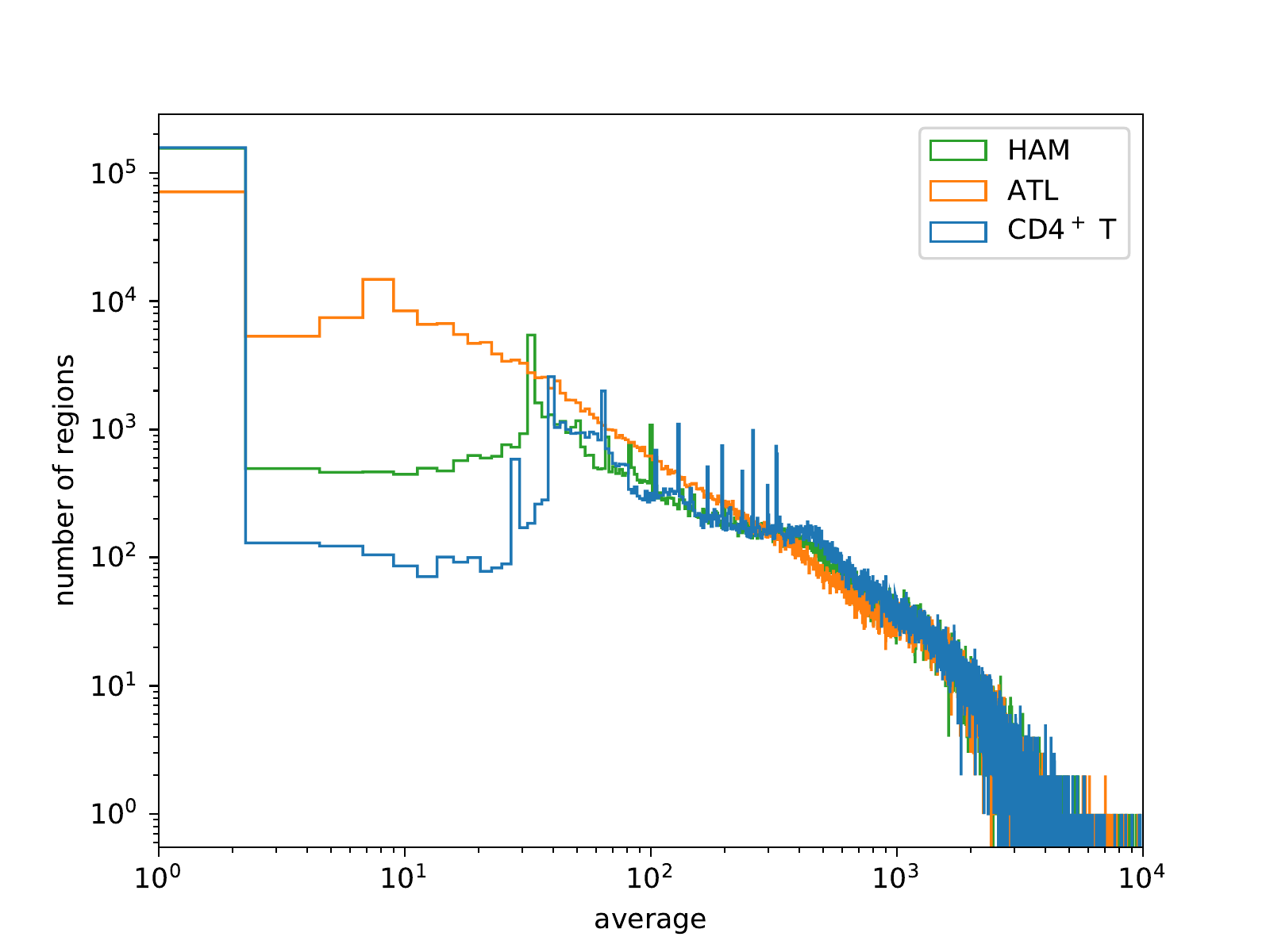}
  \end{minipage}
  \vspace{15pt}
  
  \begin{minipage}[b]{0.45\linewidth}
    \centering
    \begin{flushleft}
        {(b) Coding regions}
    \end{flushleft}
    \includegraphics[
    trim = 0 0 0 40, clip,
    keepaspectratio, width=\linewidth]{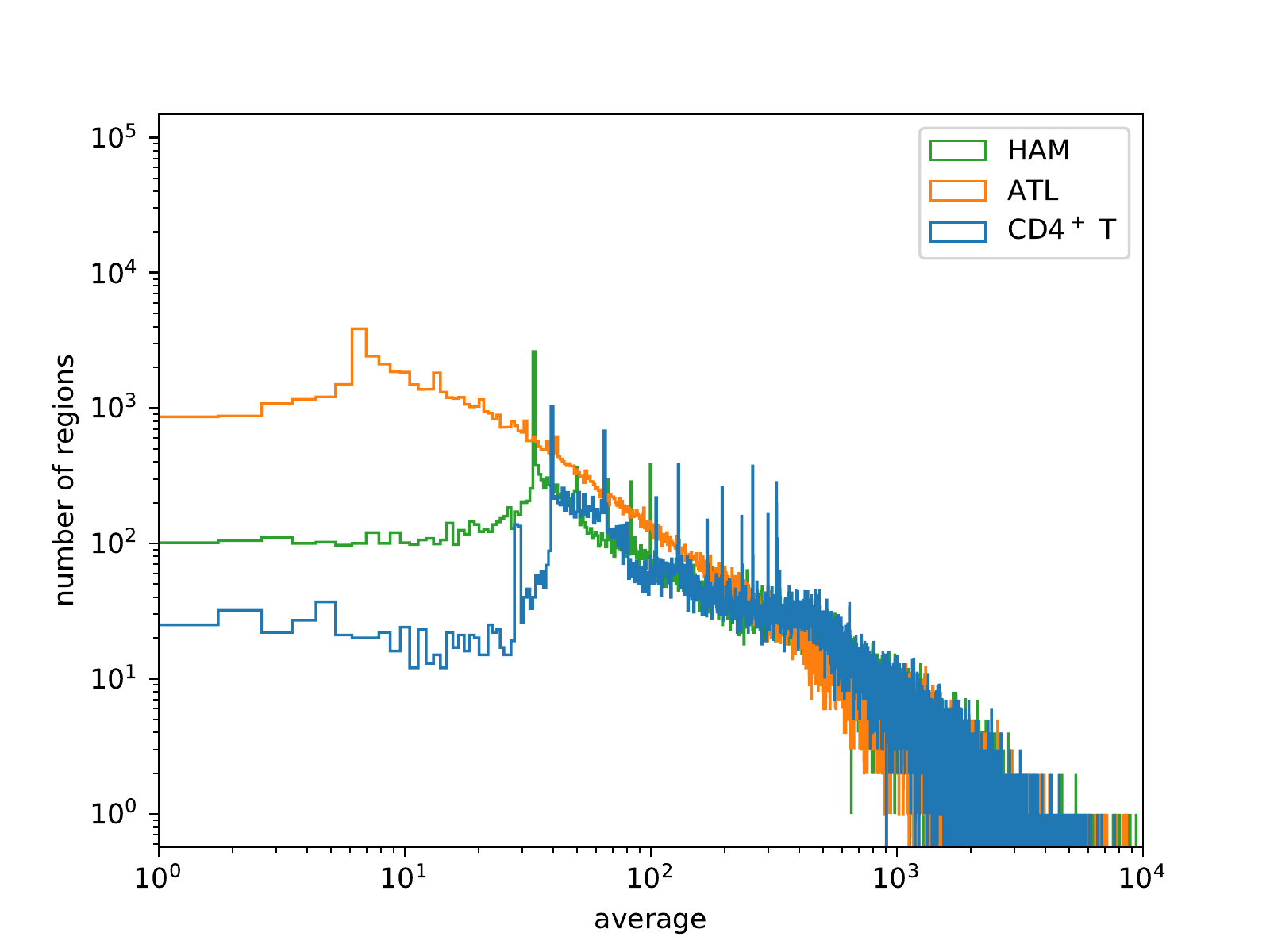}
  \end{minipage}
  \begin{minipage}[b]{0.45\linewidth}
    \centering
    \begin{flushleft}
        {(c) Non-coding regions}
    \end{flushleft}
    \includegraphics[
    trim = 0 0 0 40, clip,
    keepaspectratio, width=\linewidth]{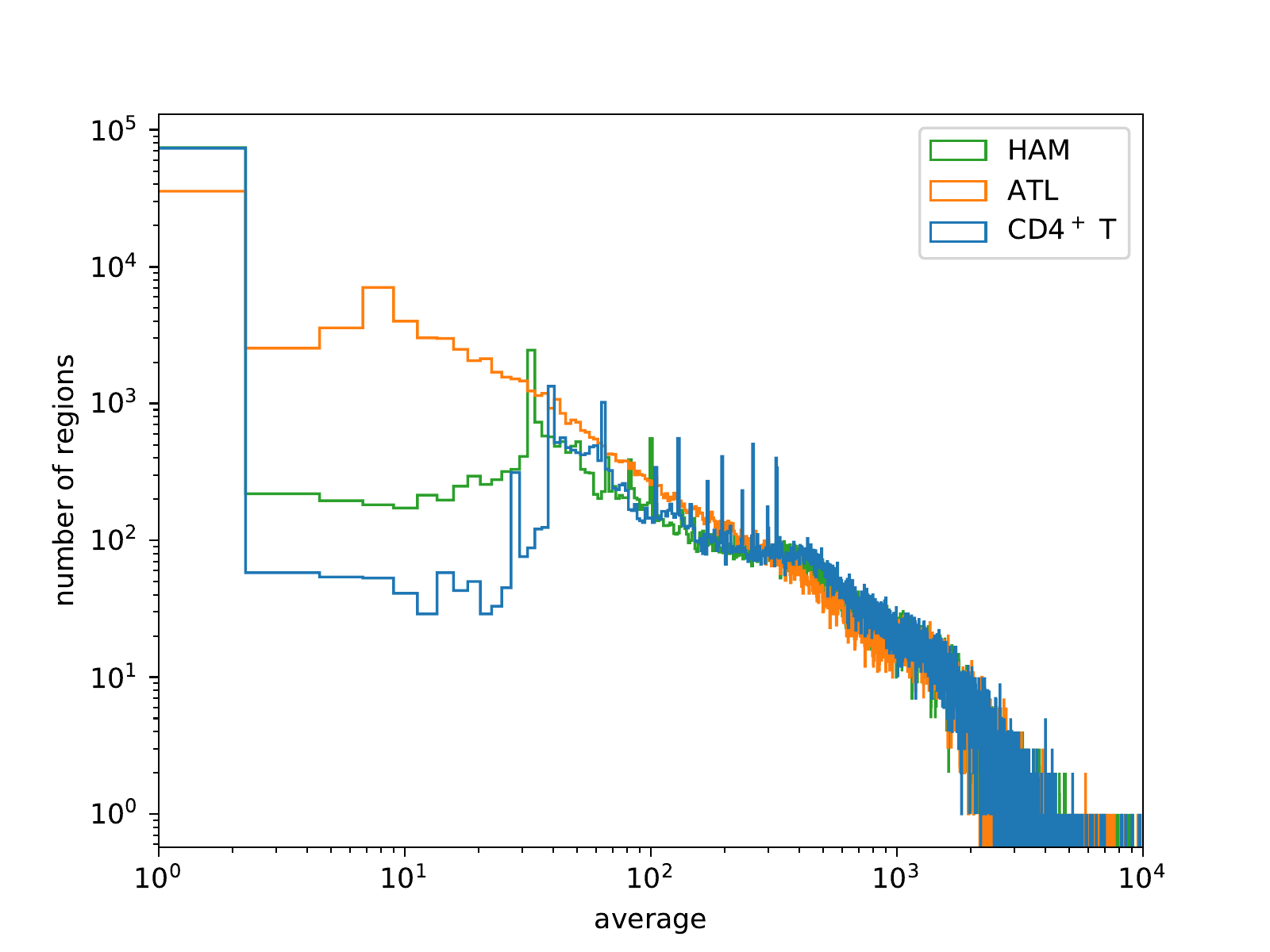}
  \end{minipage}
  \end{center}

\caption{Histograms $F^{(1)}_{\mathbb{L},\mathbb{S}}(O;\Delta)$ of the averaged length $O$ of peaks overlapping with the reference open region as defined in (\ref{eq:av_overlaps}) with bin width $\Delta$.
  $\mathbb{S}$ consists of 
  $6$ HAM samples, $29$ ATL samples, and $5$ healthy CD4$^+$T samples. (a) All regions with $\mathbb{L}=\mathbb{G}\cup\mathbb{G}^c$. (b) Coding regions with $\mathbb{L}=\mathbb{G}$. (c) Non-coding regions with $\mathbb{L}=\mathbb{G}^c$.}
\label{fig:averages}
\end{figure}

To determine how the chromatin structures observed in the HTLV1-infected cases are statistically characterized depending on the positions in the genome, we examined the positions of the reads from the ATAC-seq data. 

We plotted a histogram $\widetilde{\rho}_{\nu}(z)$ of the reads as a function of their positions $z$ relative to TSSs for cell type $\nu$, where $\nu$ is an ATL, HAM, or healthy CD4$^+$ T cell type. Note that the positions of the TSSs and coding regions of all genes were obtained from the human genome (hg19). 
For technical details of the histogram, see Materials and Methods.
As shown in Fig. \ref{fig:around_TSS}a, the tail parts of the histogram for ATL cases take higher values compared with HAM and healthy CD$4^+$ T cells, whereas the latter two cases showed more similar forms as a whole. 

To elucidate the statistics of the chromatin structures around the TSSs, we also focused on the fragments, which can be reconstructed by the reads data; both ends of a fragment correspond to a pair of two reads.  Specifically, we investigated the position-dependent accumulation of the fragments, which can be an estimate of nucleosome positioning. We plotted a heat map $F^{\Delta,\xi}_{\nu} (z,\ell)$ in which the mid-point of each fragment relative to TSSs is placed on the horizontal axis as $z$ and the length of each fragment is placed on the vertical axis as $\ell$ \citep{Henikoff2011}. For technical details of the heat map, see Materials and Methods.

As shown in Fig.\ \ref{fig:around_TSS}b, healthy CD4${^+}$ T cells and HTLV-1-infected CD4${^+}$ cells from HAM samples show a pattern of enriched nucleosome-free fragments ($\ell <100$ bp) and mono-nucleosome fragments ($\ell= 180 \sim 247$ bp) surrounding the TSSs, where the thresholds for the length of a fragment are $100$ bp, $180$ bp, and $247$ bp based on a previous study \citep{Mazumdar2015}. 
ATL samples showed less enrichment of nucleosome-free and mono-nucleosome fragments. 

These observations suggest that the statistics of open chromatin regions and nucleosome positioning around the TSSs in ATL cases is distinct from HAM cases and healthy CD4$^+$ cells, both of which showed more similar forms again. We continue to elaborate on characteristic behaviors of ATL cells distinct from the other two cases.

\subsection{Giant sample-to-sample fluctuations of chromatin accessibility in ATL} \label{Rsection2}

\begin{figure}[t]
\begin{center}

\begin{minipage}[b]{0.45\linewidth}
\centering
\begin{flushleft}
        {(a) All regions}
\end{flushleft}
\includegraphics[clip, trim = 0 0 0 40, width=\linewidth]{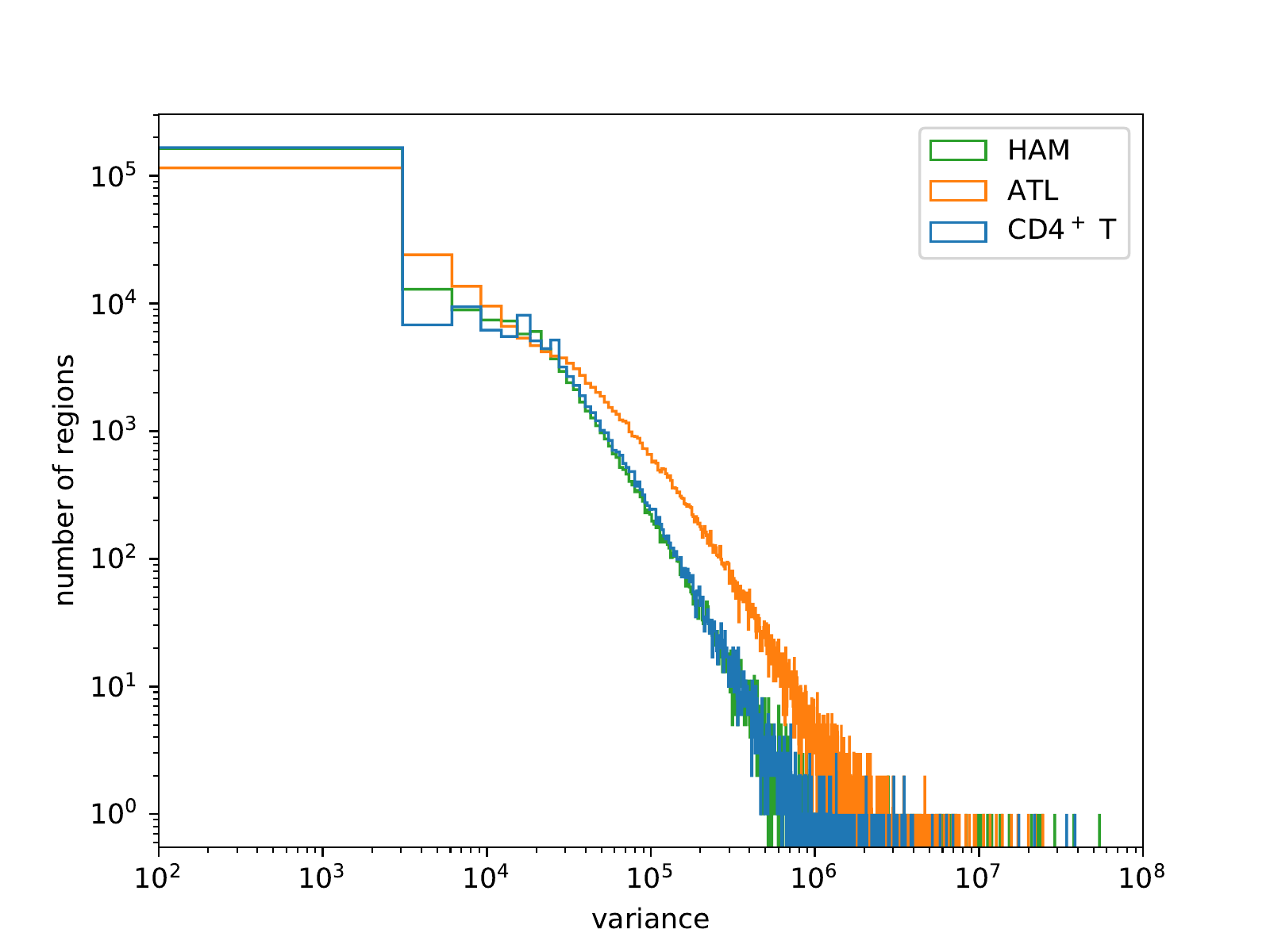}
\end{minipage}
\begin{minipage}[b]{0.45\linewidth}
\centering
\begin{flushleft}
{(b) Coding regions}
\end{flushleft}
\includegraphics[clip, trim = 0 0 0 40, width=\linewidth]{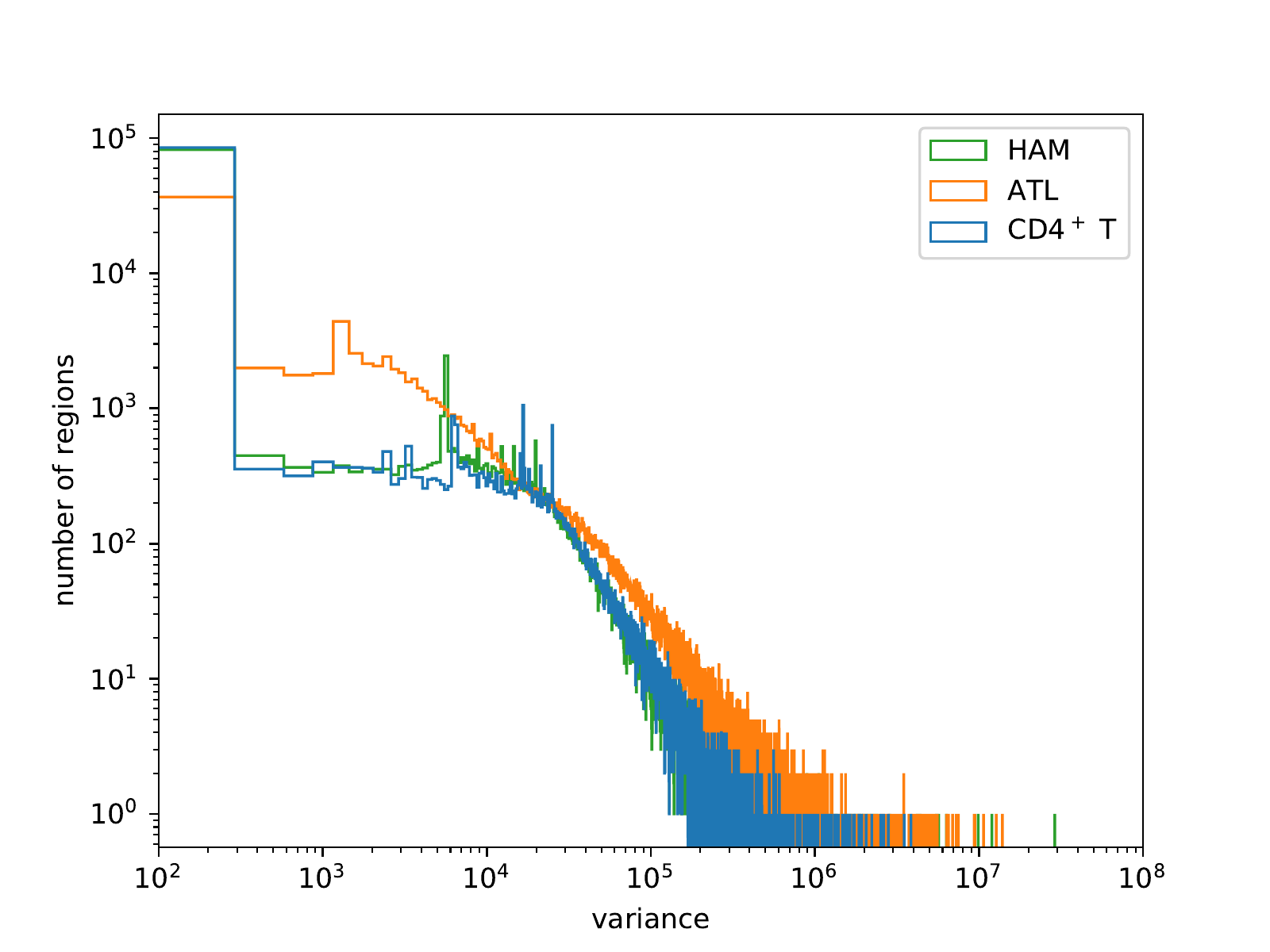}
\end{minipage}

 \vspace{15pt}

\begin{minipage}[b]{0.45\linewidth}
\centering
\begin{flushleft}
{(c) Non-coding regions}
\end{flushleft}
\includegraphics[
    trim = 0 0 0 40, clip, width=\linewidth]
{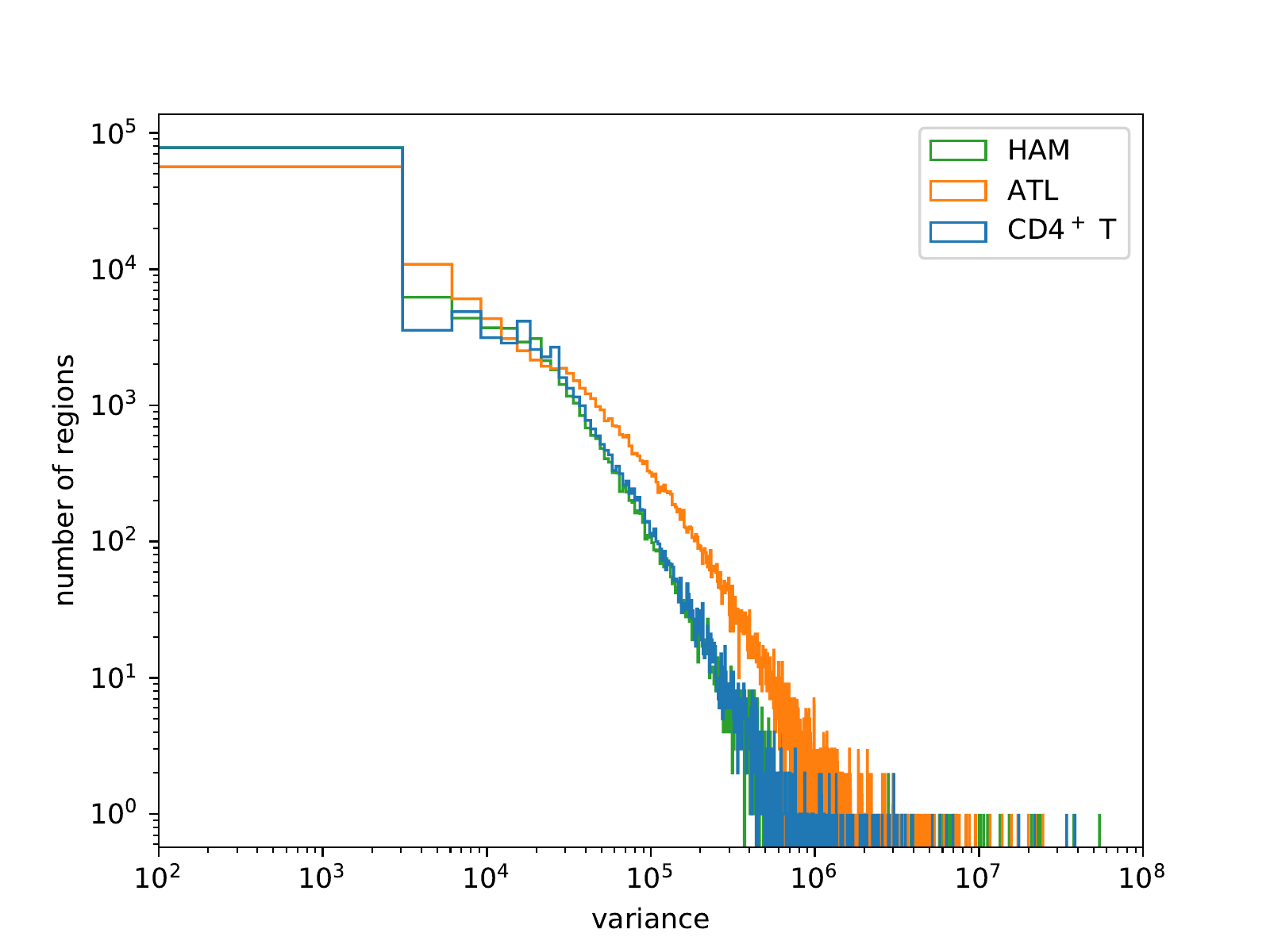} 
\end{minipage}
\begin{minipage}[b]{0.45\linewidth}
\centering
\begin{flushleft}
{(d) All regions}
\end{flushleft}
     \includegraphics[
     trim = 0 0 0 40, clip, width=\linewidth]{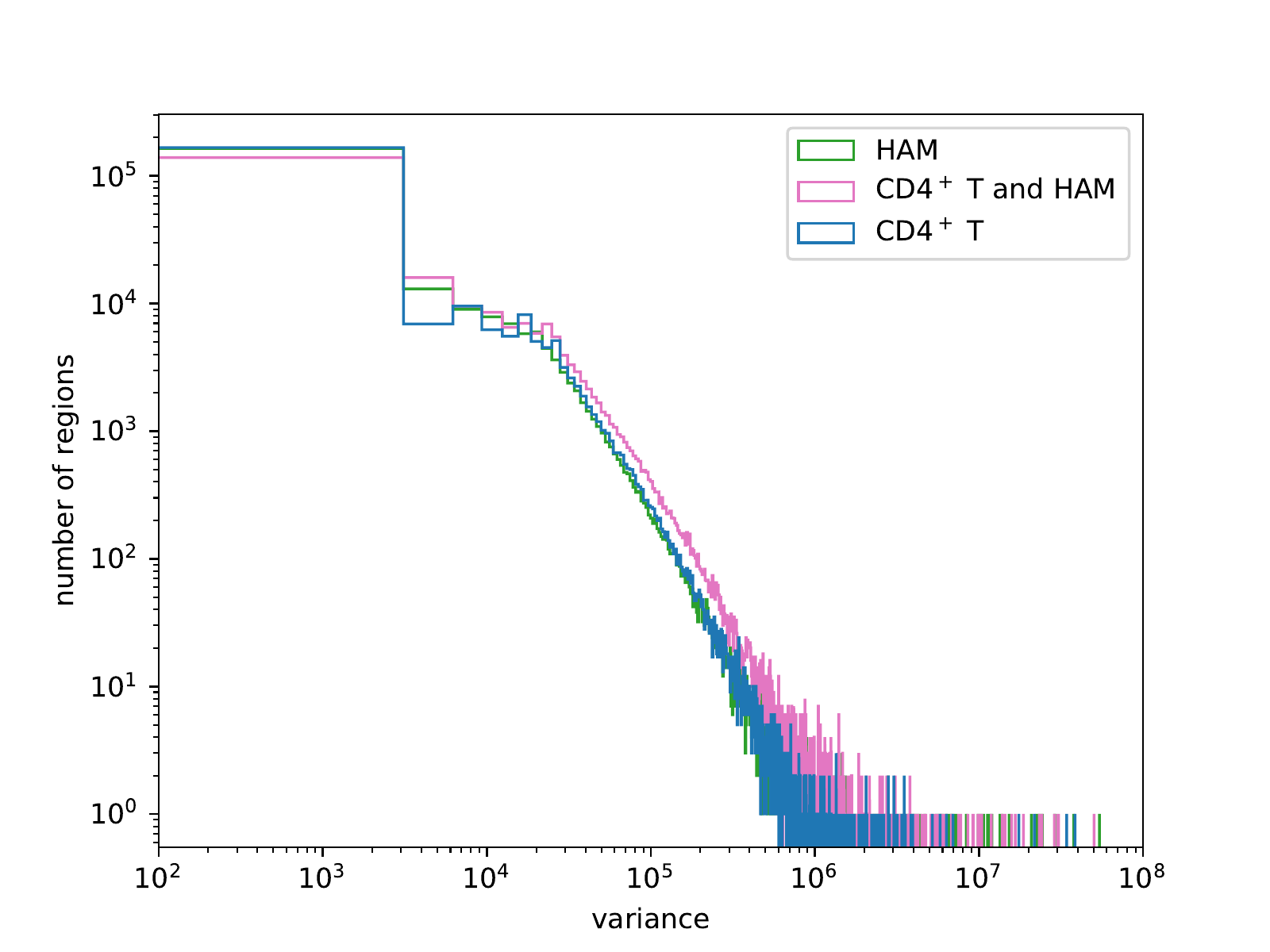}
     \end{minipage}
     
\end{center}
\caption{Histograms $F^{(2)}_{\mathbb{L},\mathbb{S}}(O;\Delta)$ of the  variance in length $O$ of peaks overlapping the reference open region as defined in (\ref{eq:var_overlaps}) with bin width $\Delta$.
  $\mathbb{S}$ consists of $6$ HAM samples, $29$ ATL samples, and $5$ CD4$^+$T samples for (a), (b), and (c).
  (a) All regions with $\mathbb{L}=\mathbb{G}\cup\mathbb{G}^c$. 
  (b) Coding regions with $\mathbb{L}=\mathbb{G}$. 
  (c) Non-coding regions with $\mathbb{L}=\mathbb{G}^c$.
  (d) All regions with $\mathbb{L}=\mathbb{G}\cup\mathbb{G}^c$ for CD4$^+$ T only, HAM only, and a mixed set of CD4$^+$ T and HAM with $5$ samples for each cell type.}
\label{fig:variances}
\end{figure}
 
To examine the distribution of open chromatin regions, we applied a systematic clustering algorithm that robustly detects open chromatin regions relevant to classifying the immunophenotype \citep{Tanaka2020}.

This algorithm characterizes open/closed chromatin regions in the following way. First, MACS2 uses ATAC-seq reads data for a given sample $s$ as the input and outputs a collection of peak data including the locations of peaks with their $p$-values.  
These peaks are considered as open regions of chromatin for sample $s$.
The peaks are ordered in ascending $p$-values, and the first $M$ peaks are taken.  The set of $M$ peaks are written as $\hat{g}_s^M = ((\gamma_k,\alpha_k,\beta_k),p_k)_{k=1}^{M}$, where the location of the $k$-th peak is the region $(\alpha_k,\beta_k)$ in the $\gamma_k$-th chromosome, and $p_k$ is the $p$-value of the $k$-th peak.
Then, the set of the top $M = 64000$ peaks is determined as the optimal set of open chromatin regions, which effectively classifies the immunophenotypes of the samples \citep{Tanaka2020}.
In this procedure, peaks with high $p$-values (unreliable peaks) can be treated as noise and ignored for later analysis.

First, we tried to capture the genomic positions where chromatins tend to be open for at least one of the ATL, HAM, and healthy CD4$^+$T types. 
To quantify such chromatin regions, we constructed a new reference set $g_0$ of peaks as follows: we concatenated all the reads from all the samples with the same cell type. Then, we used MACS2 to obtain a set of peaks for a cell type of the concatenated reads. Finally, we merged all the peaks obtained from the three cell types. For the explicit construction of the reference set $g_0$, see Materials and Methods.
Next, we classified the reference set $g_0$ into the set of all peaks overlapping gene-coding regions, which we denoted as $\mathbb{G}$, and the set of all peaks overlapping with non-coding regions, which we denoted as $\mathbb{G}^c$.

To quantify open chromatin regions characterized by the peaks $\hat{g}_s^M$ of sample $s$ in each $k$-th peak from the reference set $g_0$, we computed the length of overlapped peaks $O_k^{\mathbb{L}}(g_0, \hat{g}_s^M)$ between $g_0$ and $\hat{g}_s^M$  in the peak location $(\alpha_k,\beta_k)$ picked up from the reference region $\mathbb{L}\in\{\mathbb{G},\mathbb{G}^c,\mathbb{G}\cup \mathbb{G}^c\}$.
We set $M=64000$ as the provisionally optimal number for the immunophenotype classification \citep{Tanaka2020}. 
Then, we focused on the average and variance of $O_k^{\mathbb{L}}(g_0, \hat{g}_s^M)$.
For details of the calculations, see Materials and Methods.

As shown in Fig.\ \ref{fig:averages}, healthy CD4$^+$ T, HAM, and ATL cells showed similar behaviors at large average lengths of the overlapped peaks.
For small average lengths, CD4$^+$ T showed fewer peaks, compared with ATL cases.
Additionally, as shown in Fig.\ \ref{fig:variances}a, 
we found that healthy CD4$^+$ T and HAM cases had similar sample-to-sample fluctuations, and ATL cases had a higher frequency at variances larger than $10^5$.
 As shown in Fig.\ \ref{fig:variances}b and Fig.\ \ref{fig:variances}c, the larger sample-to-sample fluctuations in the ATL cases were found in both non-coding regions and coding regions. 
 In contrast, Fig.\ \ref{fig:variances}b shows apparent gaps between ATL cases and the other two cases at intermediate variances only in coding regions; ATL cases had a higher frequency at intermediate variances around $10^3$ compared with the other two cases.

The above analysis does not clearly distinguish healthy CD4$^+$ T and HAM cells, in particular, with respect to the sample-to-sample fluctuations of the open chromatin regions. 
Therefore, we mixed the datasets of 5 cases of HAM cells and 5 cases of healthy CD4$^+$ T cells to give set $\mathbb{S}$.
As shown in Fig.\ \ref{fig:variances}d, the variance in the mixed dataset was larger than of HAM cells or healthy CD4$^+$ T cells alone.
This finding indicates the distributions of open chromatin regions are different between healthy CD4$^+$ T and HAM cells, although the two distributions showed similar sample-dependence.
As a comparision, Fig.\ \ref{fig:variances}d and Fig.\ \ref{fig:variances}a show that the sample-to-sample fluctuations of ATL cells are larger than the fluctuations of the mixed data except for the tail, where the samples are scarce.

Thus, ATL cases have a higher frequency at larger sample-to-sample fluctuations at the whole genome level and a higher frequency at intermediate sample-to-sample fluctuations only in coding regions.
On the other hand, the chromatin structures in HAM cases show less sample-dependence, which is similar to CD4$^+$ T cells, implying the existence of a certain trend common to all the HAM samples.

\begin{figure}[t]

   \begin{minipage}[b]{0.4\linewidth}
     \centering
     \begin{flushleft}
     {(a)}
     \end{flushleft}
     \includegraphics[trim = 0 0 0 40, clip, width=\linewidth]{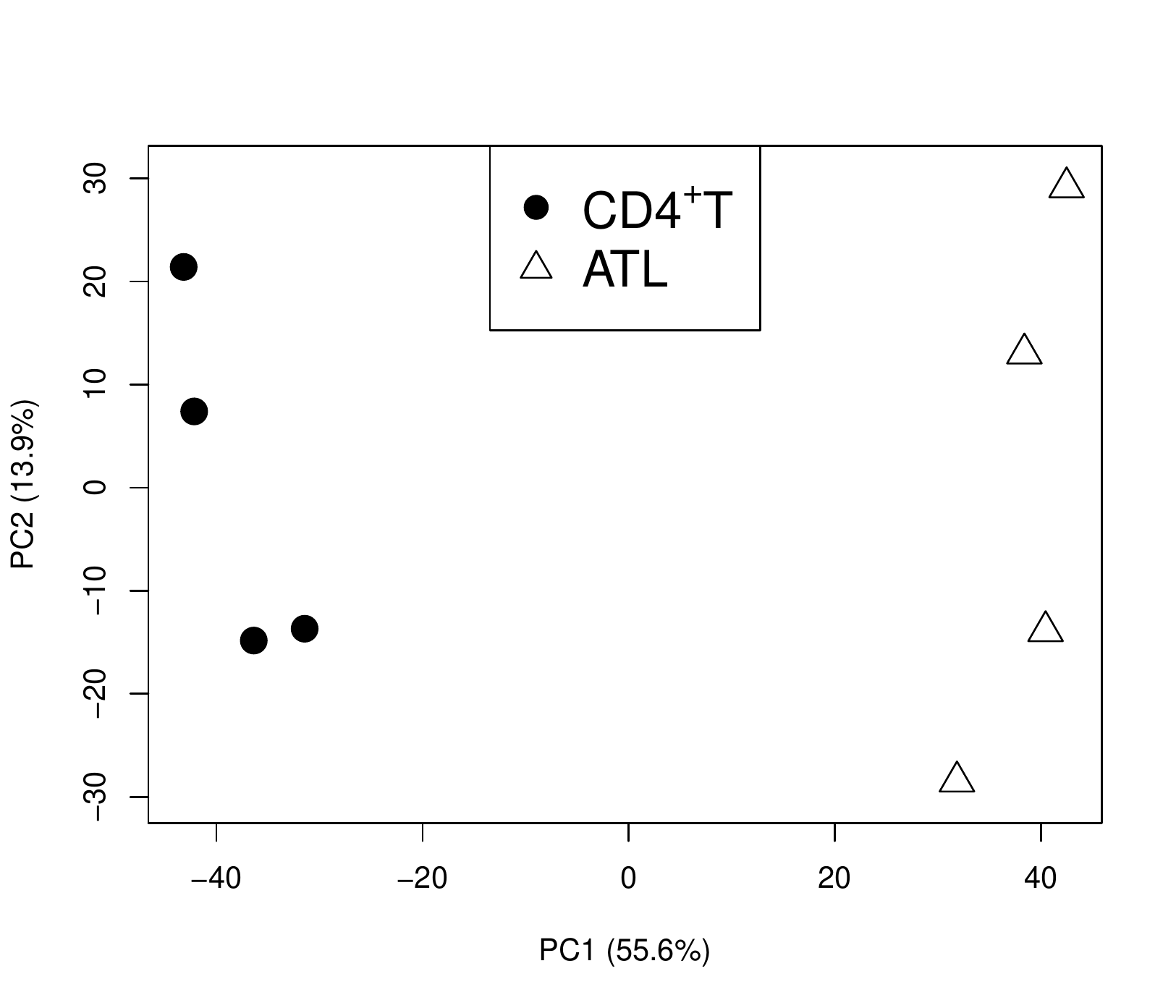}
    \begin{flushleft}
     {(c)}
     \end{flushleft}
     \includegraphics[trim = 0 0 20 30, clip, width=\linewidth]{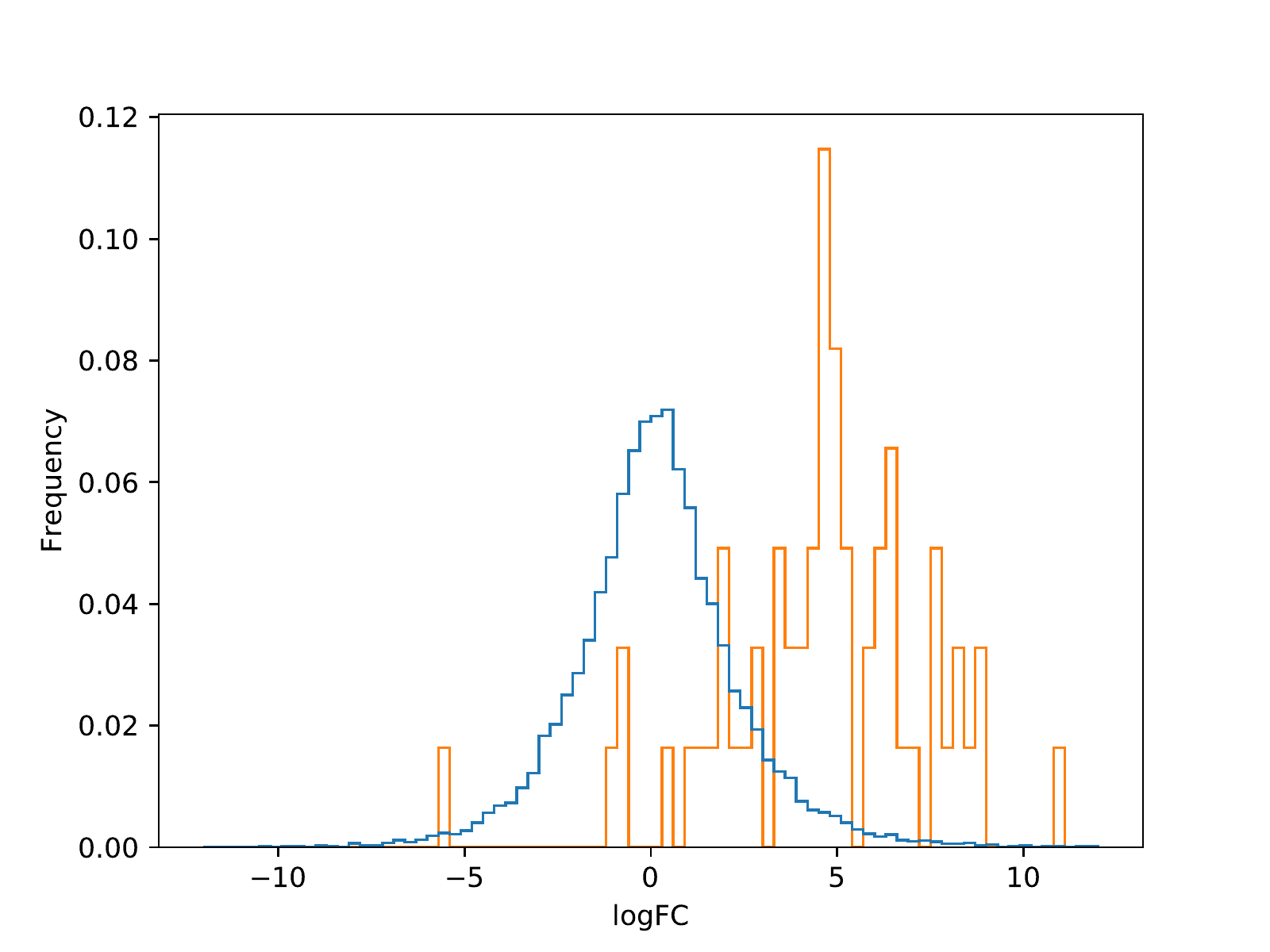}
   \end{minipage}  
 \begin{minipage}[b]{0.59\linewidth}
     \centering
    \begin{flushleft}
         {(b)}
     \end{flushleft}
     \includegraphics[clip, width=\linewidth]{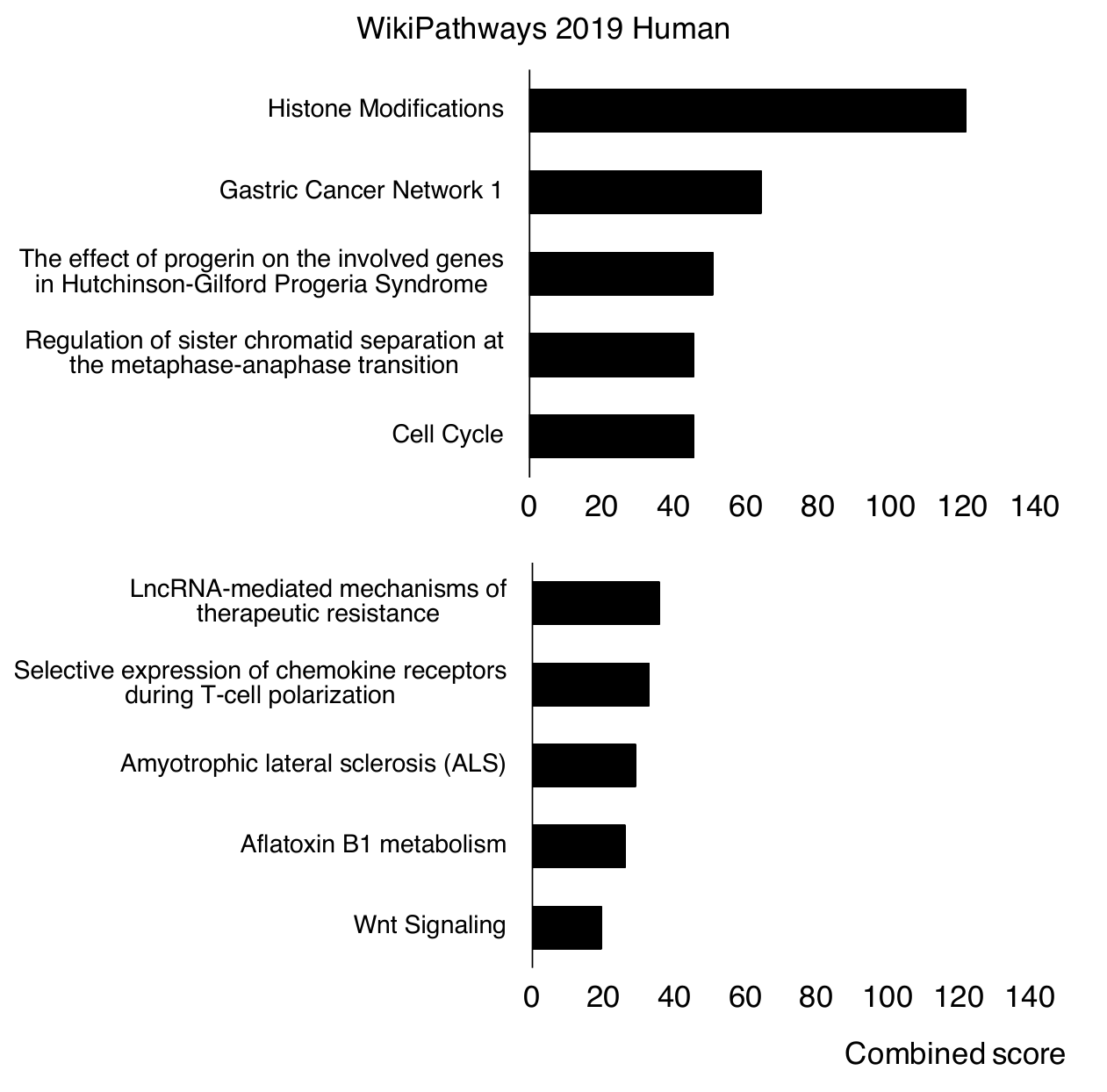}
    \end{minipage}


\caption{ RNA-seq statistics:
  (a) A PCA of RNA-seq for healthy CD4$^+$T cells and ATL cells, where the percentages are the first and second contribution ratios.
  (b) A GO analysis of RNA-seq for detecting the top 5 up-regulated gene expressions (top) and the top 5 down-regulated expressions (bottom) in terms of cell function for ATL.
  (c) A histogram of the log-fold change calculated in (\ref{eq:F}) and (\ref{eq:rho}) between the gene expressions (RNA-seq) of ATL cases and of healthy CD4$^+$ T cells for all genes (blue line) vs. 61 histone-related genes (orange line). To select relevant genes, first we chose 69 genes whose names start with HIST. Then, we removed 8 genes, including 7 genes for which healthy CD4$^+$ T cells had no peak and 1 gene for which ATL cells had no peak.
}
\label{fig:RNA-seq}
\end{figure}

\subsection{mRNA: distinctively expressed histone modifications in ATL} \label{Rsection6}

To analyze gene expressions, we examined the RNA-seq data of ATL cases. 
Unless noted otherwise, the samples used were 8, 10, 21, 24 in Table \ref{tab2}, which had common properties in terms of immunophenotypes and symptoms, as explained below.

We analyzed the gene expression pattern of HTLV-1-infected CD4$^+$ cells obtained from the peripheral blood of 4 ATL cases and of 4 healthy CD4${^+}$ T cells. The reads count of the RNA-seq data was normalized by TMM normalization \citep{cite-key} and used as the input data. For technical details, see Materials and Methods.

As shown in Fig.\ \ref{fig:RNA-seq}a, a principal component analysis (PCA) shows that the gene expression patterns differ significantly between ATL and healthy CD4$^+$ T cells. In ATL cells, there were $1289$ genes up-regulated based on the condition of $\log_2\rm{FC}>3$ and $p$-value $<0.01$ and $944$ genes down-regulated based on the condition of $\log_2\rm{FC}<-3$ and $p$-value $<0.01$, where $\mathrm{FC}$ is the fold change of the gene expression of ATL cells relative to healthy CD4${^+}$ T cells.

In addition, as shown in Fig.\ \ref{fig:RNA-seq}b, a Gene Ontology (GO) analysis using enrichR \citep{enrichr} revealed that in ATL cases, the up-regulated genes are enriched in histone modifications. Note that the combined scores of the genes with down-regulated expression are lower than those of up-regulated genes enriched for histone modification.
Further, as shown in Fig.\ \ref{fig:RNA-seq}c, many histone-related genes, such as {\it HIST1H2AH}, {\it HIST1H3C}, and {\it HIST1H4C}, are significantly up-regulated in ATL cases, where all genes beginning with HIST in the first 4 letters are regarded as histone-related genes. 
For details of the analysis, see equation \eqref{eq:F0}. 

This observation led us to consider a correlation between the anomalous chromatin properties of ATL cells shown above and the gene expression levels found here.

\begin{figure}
  \begin{center}
     
      \begin{minipage}[b]{0.48\linewidth}
     \centering
     \begin{flushleft}
    {(a) CD4$^+$T vs Mono}
    \end{flushleft}
     \includegraphics[trim = 0 0 0 30, clip, width=\linewidth]{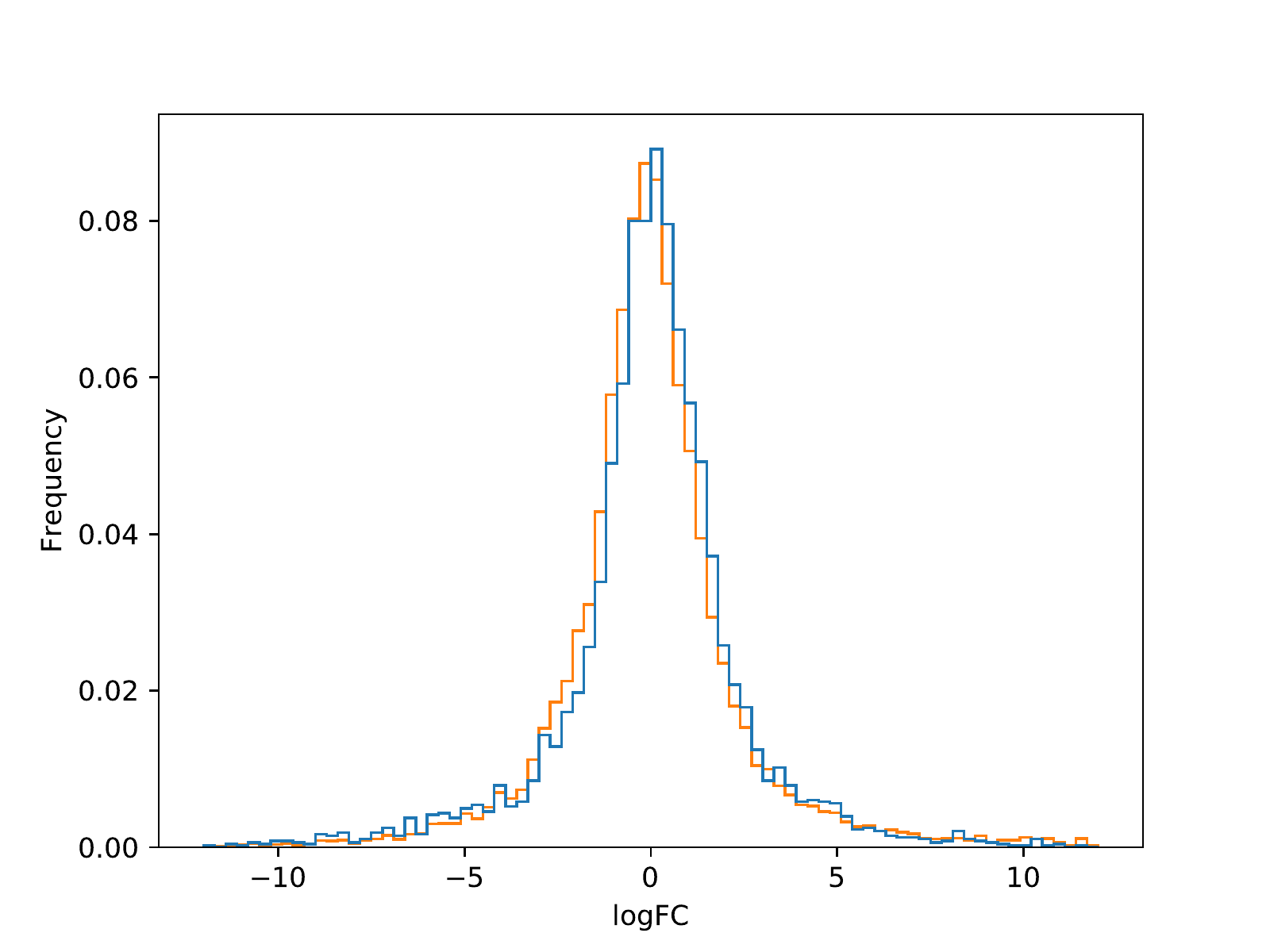}
    \end{minipage}
    \hfill
    \begin{minipage}[b]{0.48\linewidth}
      \centering
      \begin{flushleft}
      {(b) CD4$^+$T vs ATL}
      \end{flushleft}
      \includegraphics[trim = 0 0 0 30, clip, width=\linewidth]{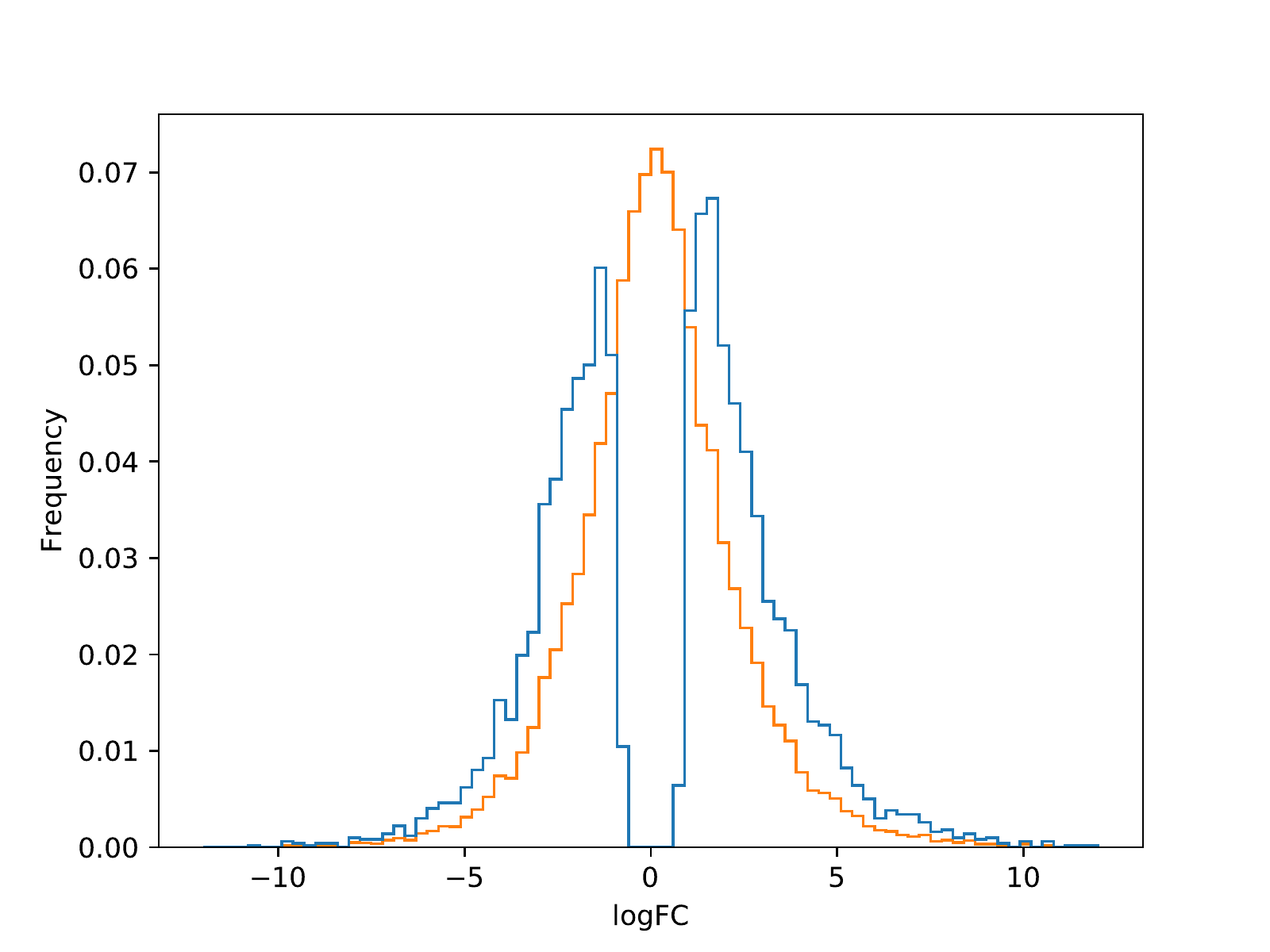}  
    \end{minipage}    
    
    \begin{minipage}[b]{0.48\linewidth}
     \centering
     \begin{flushleft}
     {(c) Mono vs ATL}
     \end{flushleft}
     \includegraphics[trim = 0 0 0 30, clip, width=\linewidth]{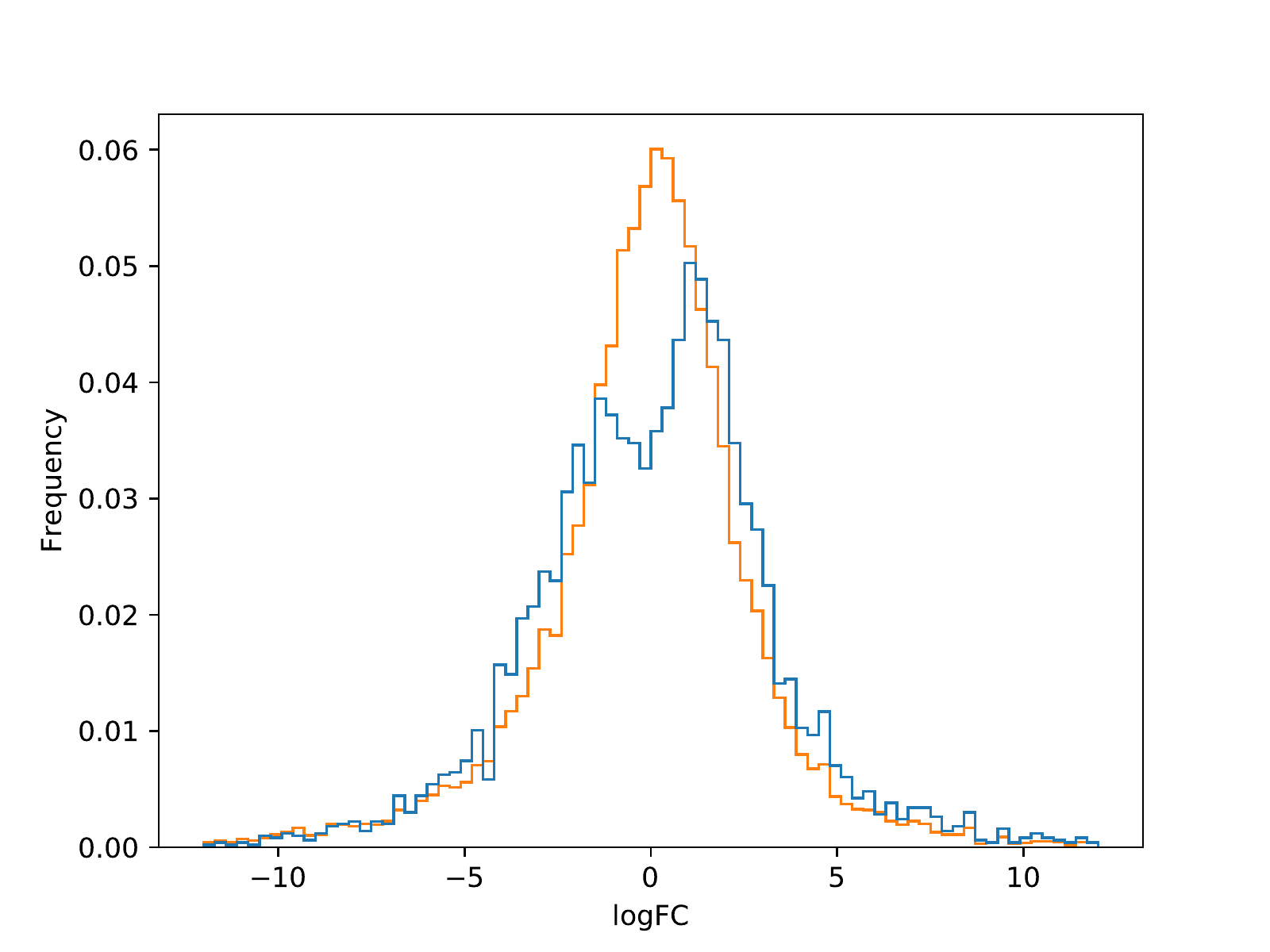}
    \end{minipage} 
    \hfill
    \begin{minipage}[b]{0.48\linewidth}
     \centering
     \begin{flushleft}
     {(d) CD4$^+$T vs CD8$^+$T}
     \end{flushleft}
     \includegraphics[trim = 0 0 0 30, clip, width=\linewidth]{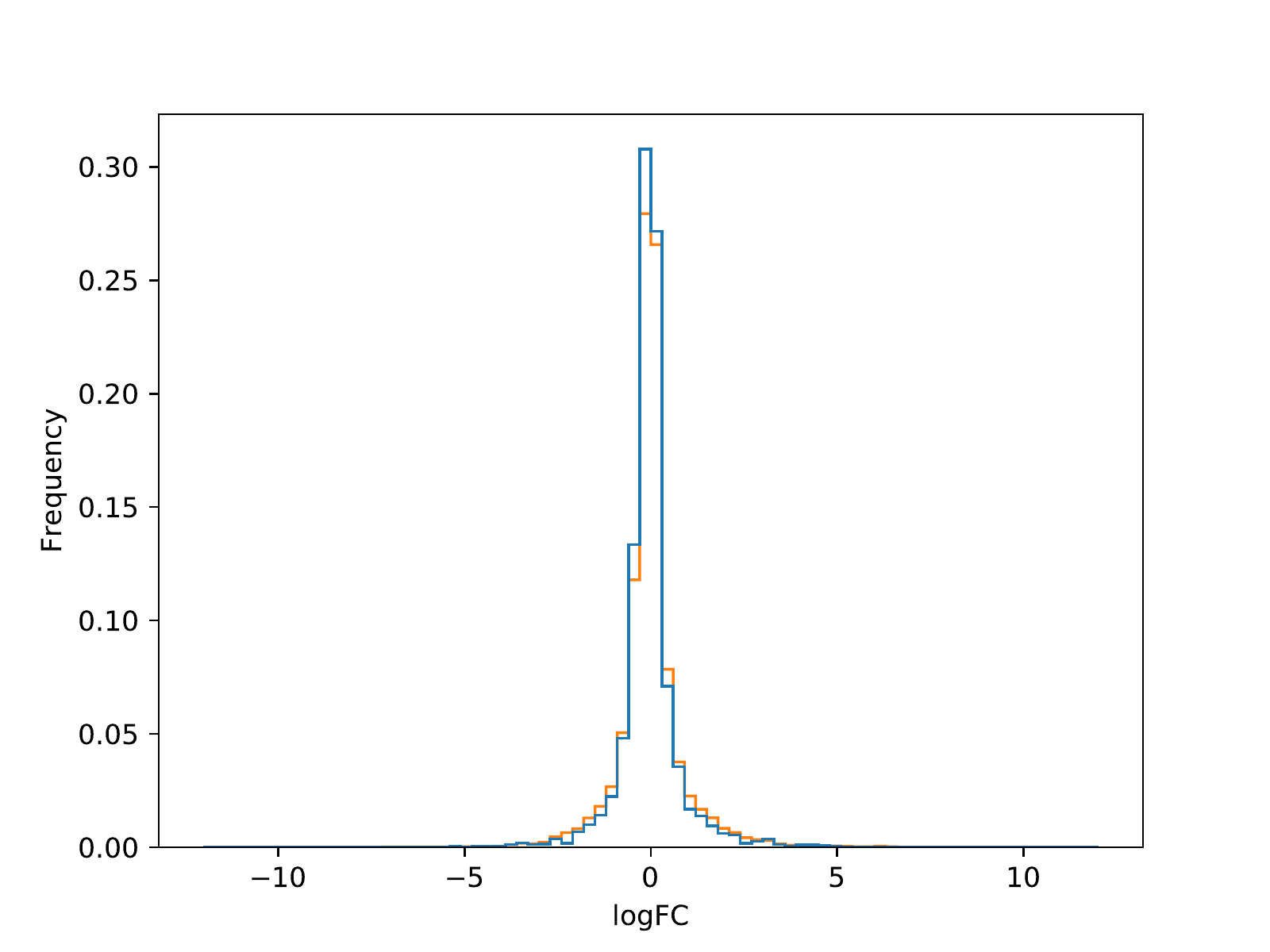}
    \end{minipage}
    
    \begin{minipage}[b]{0.48\linewidth}
     \centering
     \begin{flushleft}
     {(e) HSC vs ATL}
     \end{flushleft}
     \includegraphics[trim = 0 0 0 30, clip, width=\linewidth]{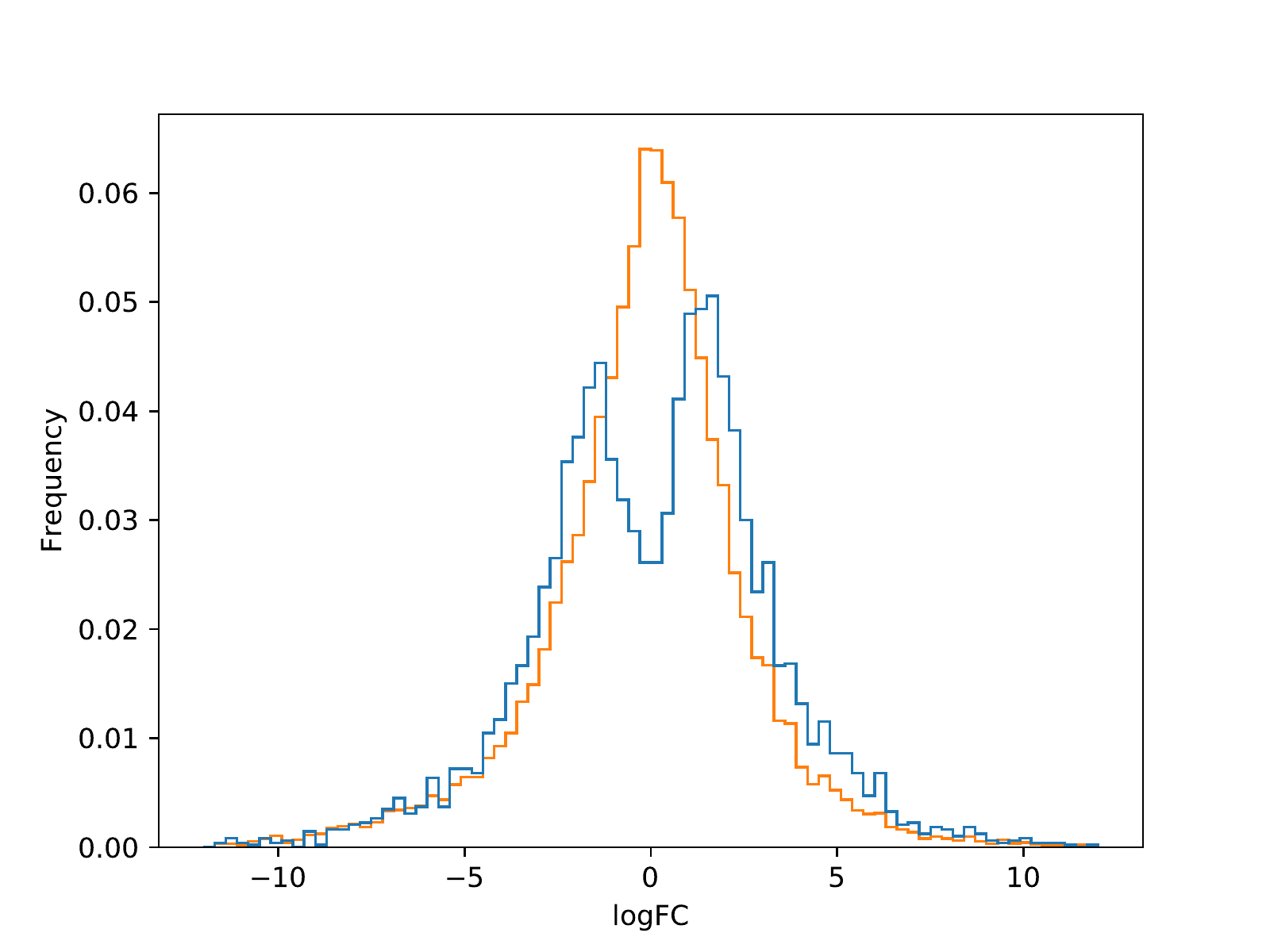}
    \end{minipage}    
    \hfill
    \begin{minipage}[b]{0.48\linewidth}
     \centering
     \begin{flushleft}
     {(f) HSC vs CD4$^+$T}
     \end{flushleft}
     \includegraphics[trim = 0 0 0 30, clip, width=\linewidth]{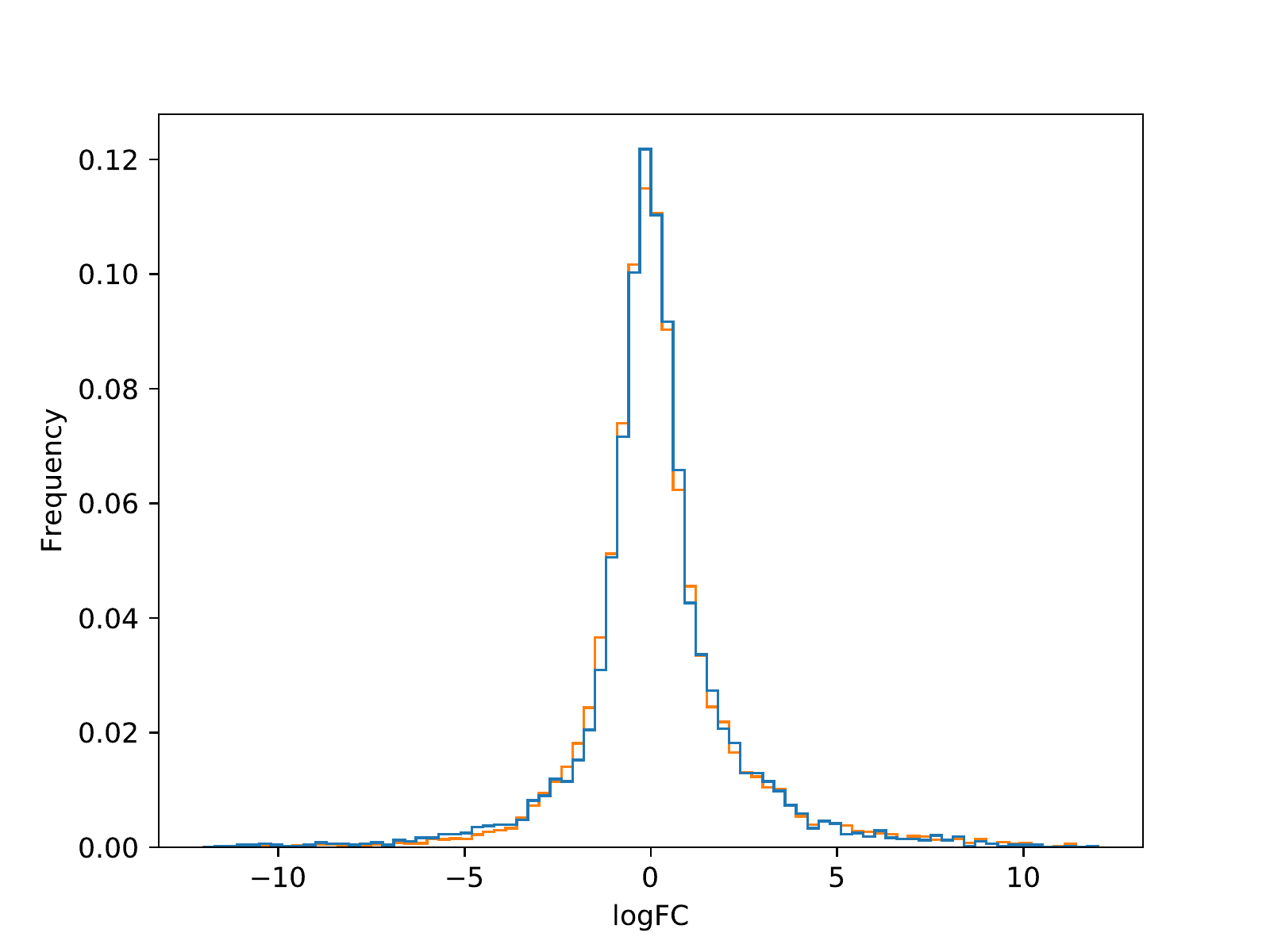}
    \end{minipage}       
  \end{center}
 \caption{
  Histograms $F_{t,t_0}^M(P;\Delta)$ of the expression of the selected genes defined in (\ref{eq:logFC}), where bin width $\Delta$ is equal to $0.3$. The histograms characterize the correlation between gene expressions quantified by RNA-seq and ATAC-seq peaks selected systematically.
  (a) $M=\infty$ (orange line), $\ 64000$ (blue line) for cell types with $(t,t_0)=(\rm{CD4}^+ \rm{T}, \rm{Mono})$.
  (b) $M=\infty,\ 64000$  for cell types with $(t,t_0)=(\rm{CD4}^+ \rm{T}, \rm{ATL})$.
  (c) $M=\infty,\ 64000$  for cell types with $(t,t_0)=(\rm{Mono}, \rm{ATL})$.
   (d) $M=\infty,\ 64000$  for cell types with $(t,t_0)=(\rm{CD4}^+ \rm{T}, \rm{CD8}^+ \rm{T})$.
  (e) $M=\infty,\ 64000$  for cell types with $(t,t_0)=(\rm{HSC}, \rm{ATL})$.
  (f) $M=\infty,\ 64000$  for cell types with $(t,t_0)=(\rm{HSC}, \rm{CD4}^+ \rm{T})$.
}
\label{fig:logFCs}
\end{figure}

\subsection{Correlation between chromatin accessibility and mRNA: exclusive mRNA expressions in ATL} \label{Rsection5}

To gain a direct quantification of how chromatin structures are correlated with gene expressions, we performed an integrated analysis of ATAC-seq and RNA-seq data for ATL cells. 

We used our algorithm to classify the top $M$ peaks into open chromatin regions, where $M$ is originally determined as $M=64000$, such that the clustering of the ATAC-seq samples is closest to the appropriate immunophenotype \citep{Tanaka2020}. 
For a comparison, we also used all peaks outputted from MACS2 as $M=\infty$.
We tried to find genes for which at least one peak from a given set of all peaks is assigned.
Note that each peak can be assigned to zero or more genes. 

Here, the RNA-seq data analyzed by edgeR \citep{Robinson2010,McCarthy2012} was used to examine the expression of each gene in ATL cells and healthy CD4$^+$T cells. 
For details of the calculations, see Material and Methods.

We considered the set of cell types $\mathbb{T}$ as
\begin{align}
 \mathbb{T}= \{\rm{HSC}, \rm{CD4}^+ \rm{T}, \rm{CD8}^+ \rm{T}, \rm{NK}, \rm{Mono}, \rm{ATL}\},\nonumber
  \end{align} 
where HSC, NK, and Mono are 
  hematopoietic stem cell, natural killer cell, and monocyte, respectively.
We computed the fold change ${\rm FC}_i(t,t_0)$ of the expression of gene $i$ between type $t,t_0\in\mathbb{T}$ as
\begin{align}\label{eq:F0}
  {\rm FC}_{i}(t,t_0):= \dfrac{\overline{R}_i(t)}{\overline{R}_i(t_0)},
\end{align} where $\overline{R}_i(t)$ means the average of the normalized reads count of the RNA-seq data in gene $i$
over all samples with type $t$.
Then, we focused on the log fold change of the gene expression $P=\log_2{\rm FC}_{i}(t,t_0)$, where we only take into account the expression of gene $i$ in which at least one peak within the top $M$ peaks from $\hat{g}_s^M$ was located for types $t,t_0$ of the samples. 
For details of the calculations, see Material and Methods. 

As shown in Fig.\ \ref{fig:logFCs}, when $M=\infty$, the distribution is close to unimodal for almost all pairs of cell types.
When $M=64000$, the distributions significantly depend on the pairs of cell types. For example, as shown in Fig. \ref{fig:logFCs}b, \ref{fig:logFCs}c, and \ref{fig:logFCs}e, 
the distributions related to ATL cases are sharply bimodal-like.
In particular, the pair ATL and CD4$^+$ T cells was found to be distinct (Fig.\ \ref{fig:logFCs}b); the events around $P=0$ are completely undetected. 

Thus, the above observations suggest that ATL cells have exceptionally distinct structures in terms of the correlation between chromatin accessibility and mRNA compared with CD4$^+$ T cells.


\begin{figure}[t]
\begin{center}
   \includegraphics[clip, width=\linewidth]{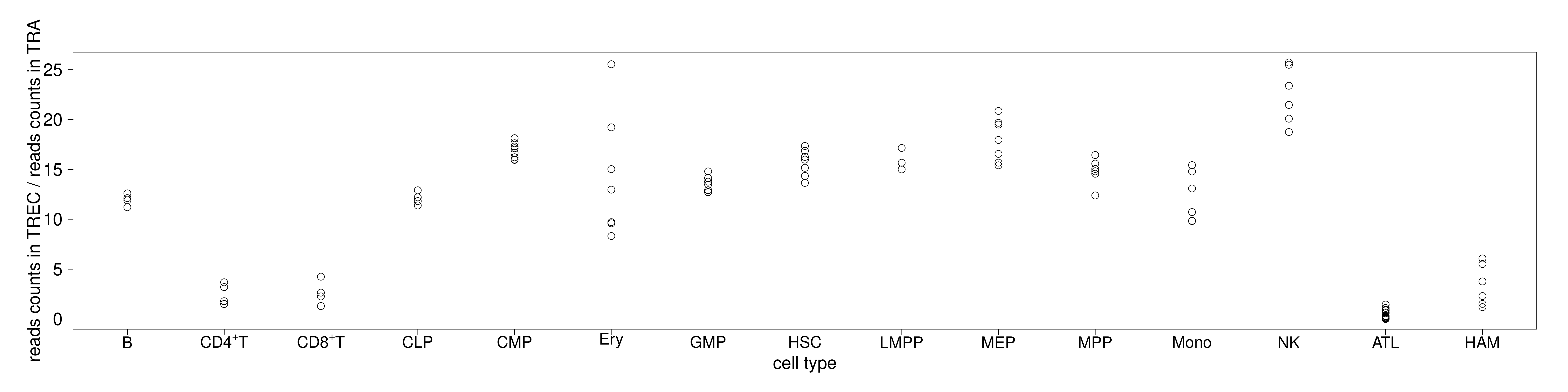}
\end{center}
\caption{Reads counts overlapping with the TREC region divided by reads counts overlapping with the TRA gene region.}
\label{fig:TREC}
\end{figure}

\subsection{Immunophenotypes from chromatin accessibility and mRNA} \label{Rsection9}

\begin{table}[hbtp]
\caption{Clustering results of ATAC-seq from ATL cases in terms of immunophenotypes obtained using the method in \citep{Tanaka2020}.
ATAC-seq information for ATL samples is DRR250714 for ATL8, DRR250710 for ATL10, DRR250711 for ATL21, DRR250712 for ATL24, DRR250713 for ATL4, DRR250715 for ATL2, and DRR250716 for ATL5. Other information about the sample labels will be added in next version.}
\label{tab2}
 \begin{tabular}{|r|c|c|c|c|}
\hline
Sample  & first & second  & third & Clinical  \\
labels &  &   &  & subtype  \\
\hline   
1 & Ery & Mono & CLP & Acute \\
2 & Mono & CD4$^+$ T & B & Chronic \\
3 & CD4$^+$ T  & CD8$^+$ T  & NK & Chronic \\
4 & CD4$^+$ T & CD8$^+$ T & B & Acute \\
5 & Mono & Ery & CD4$^+$ T & Chronic   \\
6 & Ery & Mono & B & Chronic   \\
7 & CD4$^+$ T & CD8$^+$ T & B & Acute \\
8 & CD4$^+$ T & CD8$^+$ T & NK & Acute   \\
9 & CD4$^+$ T & CD8$^+$ T & B & Acute \\
10 & CD4$^+$ T & CD8$^+$ T & B & Acute \\
11 & CD4$^+$ T & CD8$^+$ T & B & Acute \\
12 & CD4$^+$ T & CD8$^+$ T & B & Acute \\
13 & CD4$^+$ T & CD8$^+$ T & B & Acute \\
14 & CD4$^+$ T & CD8$^+$ T & B & Acute \\
15 & CD4$^+$ T & B &  CD8$^+$ T & Acute  \\
16 & CD4$^+$ T & CD8$^+$ T & NK & Acute  \\
17 & Ery & Mono & B & Chronic \\
18 & CD4$^+$ T & CD8$^+$ T & B & Acute \\
19 & CD4$^+$ T & B & CD8$^+$ T & Acute \\
20 & CD4$^+$ T & CD8$^+$ T & B &  Chronic  \\
21 & CD4$^+$ T & B & CD8$^+$ T & Acute \\
22 & CD4$^+$ T & B & CD8$^+$ T & Acute \\
23 & Ery & CD4$^+$ T & Mono & Acute \\
24 & CD4$^+$ T & CD8$^+$ T & B & Acute \\
25 & CD4$^+$ T & CD8$^+$ T & B & Acute \\
26 & CD4$^+$ T & CD8$^+$ T & B & Acute \\
27 & CD4$^+$ T & CD8$^+$ T & B & Chronic \\
28 & CD4$^+$ T & CD8$^+$ T & B & Acute \\
29 & CD4$^+$ T & CD8$^+$ T & B & Acute \\
\hline
\end{tabular}
\end{table}

We next examined ATL cases by inferring the past cell status before infection with HTLV-1 and the current cell status in terms of immunophenotypes compared with normal hematopoietic cells.

It remains unclear why most ATL cells have an immunophenotype similar to CD4$^+$ memory T cells even though HTLV-1 infects multiple hematopoietic cells \citep{Furuta2017}. 
Specifically, by focusing on the T-cell Receptor Excision Circles (TREC) region, we computed the number of reads mapped to the TREC region using ATAC-seq data from ATL cells and $13$ human primary blood cell types from healthy donors. As shown in Fig. \ref{fig:TREC}, the number of mapped reads in the TREC region in ATL/HAM cells is closer to that of CD4${^+}$ T and CD8${^+}$ T cells compared with other hematopoietic cell types. This finding indicates that cells before infection had already differentiated into T cells.
Moreover, when combined with our findings about chromatin levels above, the results suggest a strong sample-dependent opening of the chromatin occurs in infected CD4${^+}$ T cells during the development of ATL. 

As mentioned above, the overall chromatin structures vary greatly among ATL cases and are correlated with the gene expressions. 
This phenomenon presumably affects the immunophenotype.
To investigate the effect of large-scale variations of chromatin structures on the immunophenotype, we used our algorithm to quantitatively evaluate differences between ATL cells and hematopoietic cells from healthy donors \citep{Tanaka2020}.

Using this algorithm, we calculated the Hamming distances between the peak-based binarized genome of ATL cells and of hematopoietic cells from healthy donors. 
For this purpose, we used ATL samples and $77$ ATAC-seq datasets from $13$ human primary blood cell types.
As summarized in Table \ref{tab2}, the majority of ATL samples are close to CD4$^+$ T cells, as expected by the above analysis about the past cell status.
We also found that the ATAC-seq patterns of some ATL cases are close to myeloid cells such as erythroid cells and monocytes.

To ascertain whether the mRNA expression in the ATL cells reflects the characteristics of myeloid cells, we analyzed the RNA-seq data from healthy CD4${^+}$ T cells and HTLV-1-infected CD4${^+}$ cells from 4 ATL cases (samples 8, 10, 21, 24 in Table \ref{tab2}).
We used the condition $\log_2\rm{FC} > 1$ and $p$-value $< 0.01$ to identify up-regulated genes and found two candidates: CD71 (TFRC), which is ubiquitously expressed by erythroid precursors \citep{Dong2011, Marsee2010}, and KLF4, which is highly expressed in myeloid cells and essential for monocyte differentiation \citep{Feinberg2007}. 

\subsection{Chromatin-based systematic selection of key genes in ATL cases} \label{Rsection7}

\begin{figure}[t]
\begin{center}
  \begin{minipage}[b]{0.45\linewidth}
     \centering
     \begin{flushleft}
     {(a)}
     \end{flushleft}
  \includegraphics[clip, width=7cm]{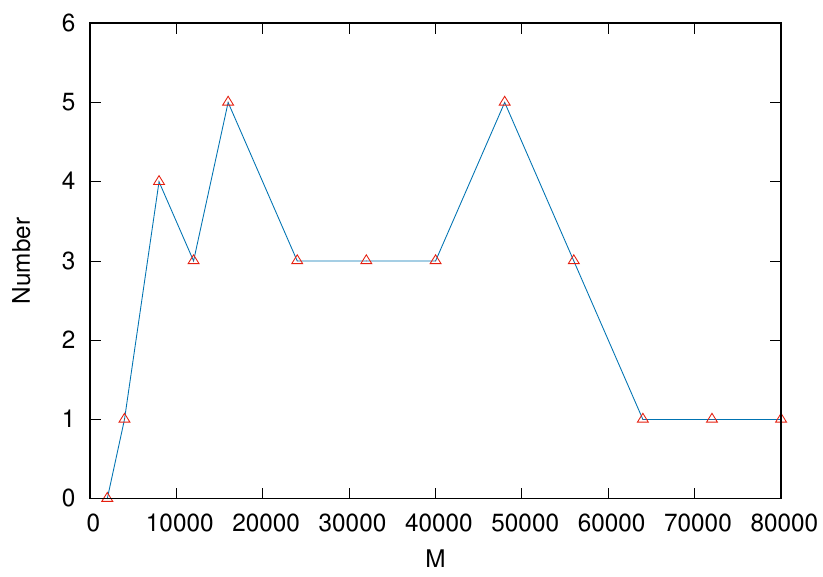}
  \end{minipage}
  \begin{minipage}[b]{0.45\linewidth}
     \centering
     \begin{flushleft}
     {(b)}
     \end{flushleft}
  \includegraphics[clip, width=7cm]{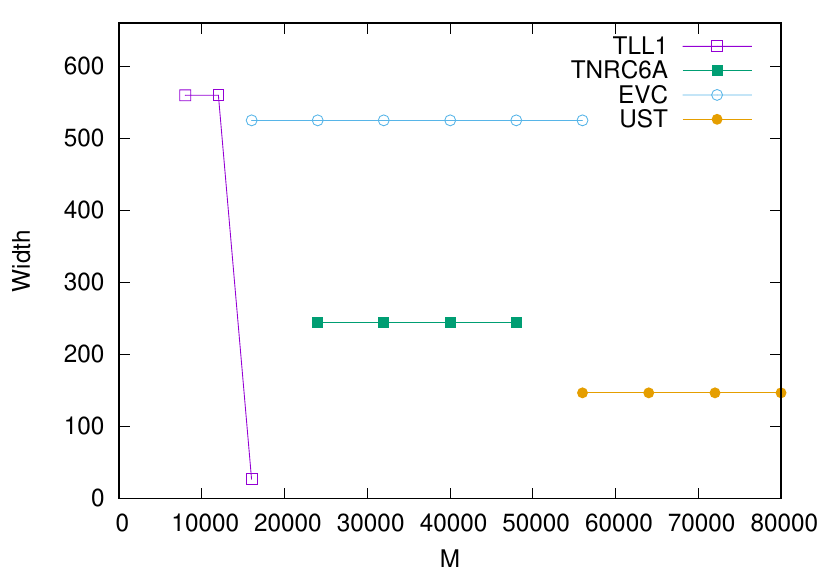}
  \end{minipage}
\end{center}
\caption{ 
  (a) The number of selected genes over increments of $8000$ for $M$ 
  from $8000$ to $80000$. (b) The width of the selected genes.}
\label{fig:selected_gene}
\end{figure}

To identify genes that are specially expressed in ATL cells, we investigated whether there are characteristic open chromatin regions in gene coding regions that are common to all ATL cells but not to hematopoietic cells derived from healthy individuals.

We therefore compared the chromatin accessibility between $29$ ATL samples and $77$ ATAC-seq datasets from $13$ human primary blood cell types.
We applied our algorithm to detect such key genes, using $M$ as a parameter for clustering the ATAC-seq data corresponding to immunophenotypes \citep{Tanaka2020}.
First, we defined a subset of regions of top $M$ peaks, where all ATL samples had peaks but no $13$ human primary blood cell types from healthy donors did. We call this subset ATL-specific open regions.
  Second, we detected the genes in which at least one of ATL-specific open region was located for a given $M$. 

\begin{figure}[t]
\begin{minipage}[b]{0.45\linewidth}
     \centering
     \begin{flushleft}
        {(a)}
    \end{flushleft}
     \includegraphics[clip, width=\textwidth]{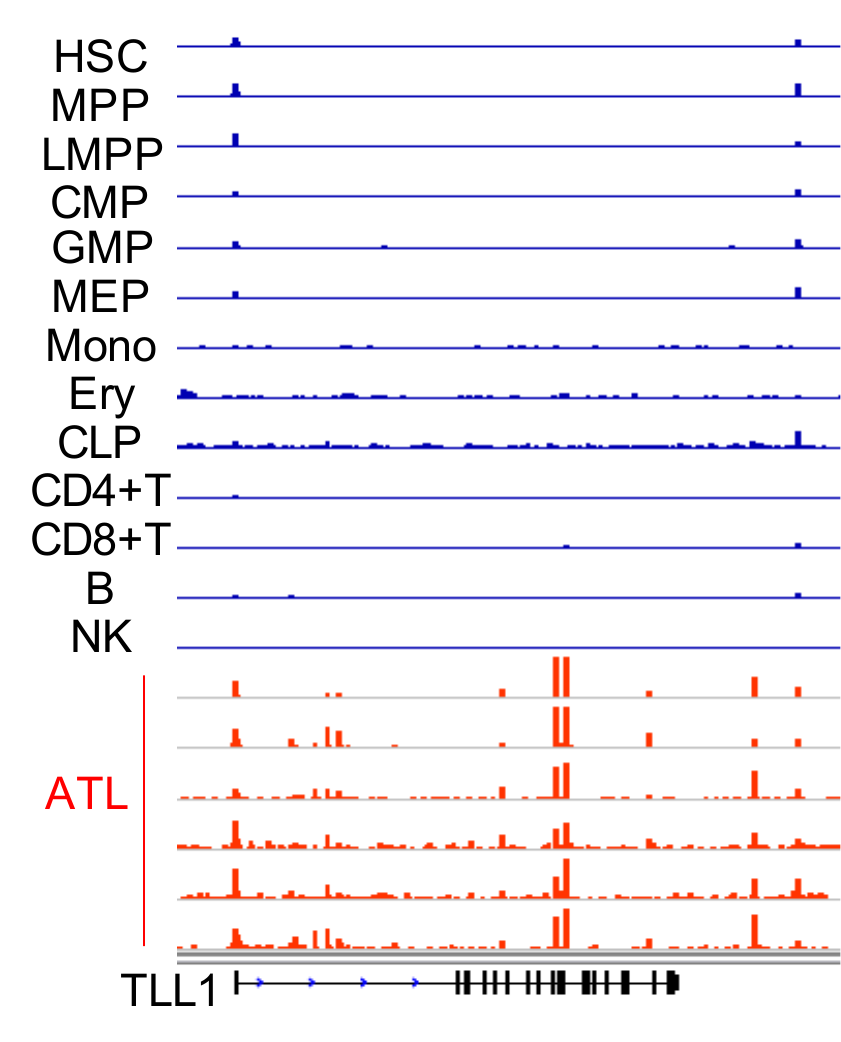}
 \end{minipage}
 \hspace{0.05\linewidth}
 \begin{minipage}[b]{0.45\linewidth}
    \begin{flushleft}
        (b)
    \end{flushleft}
    \centering
    \includegraphics[clip, width=\textwidth]{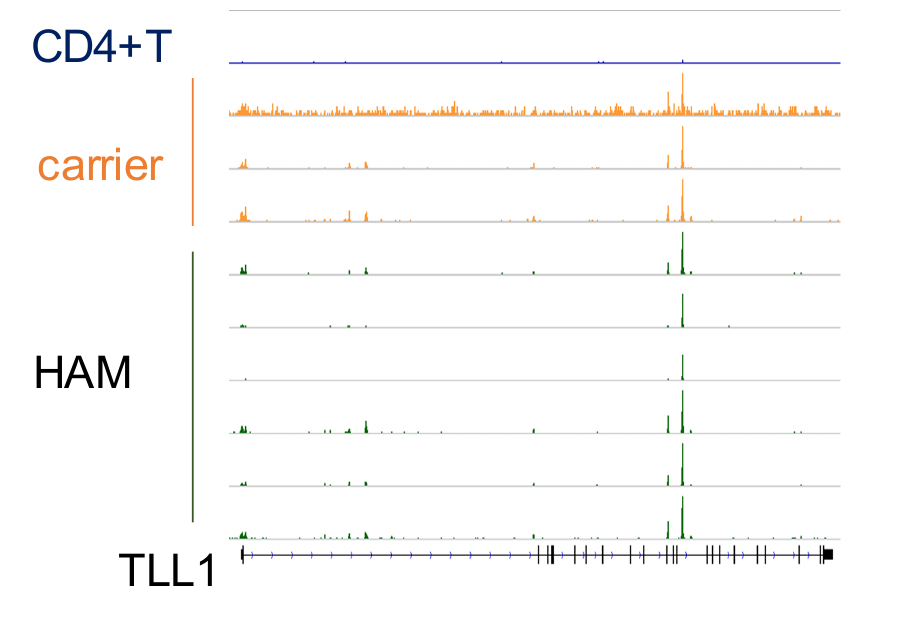}
     
    \begin{flushleft}
        (c)
    \end{flushleft}
    \includegraphics[clip, trim = 20 0 40 20, 
    width=\textwidth]{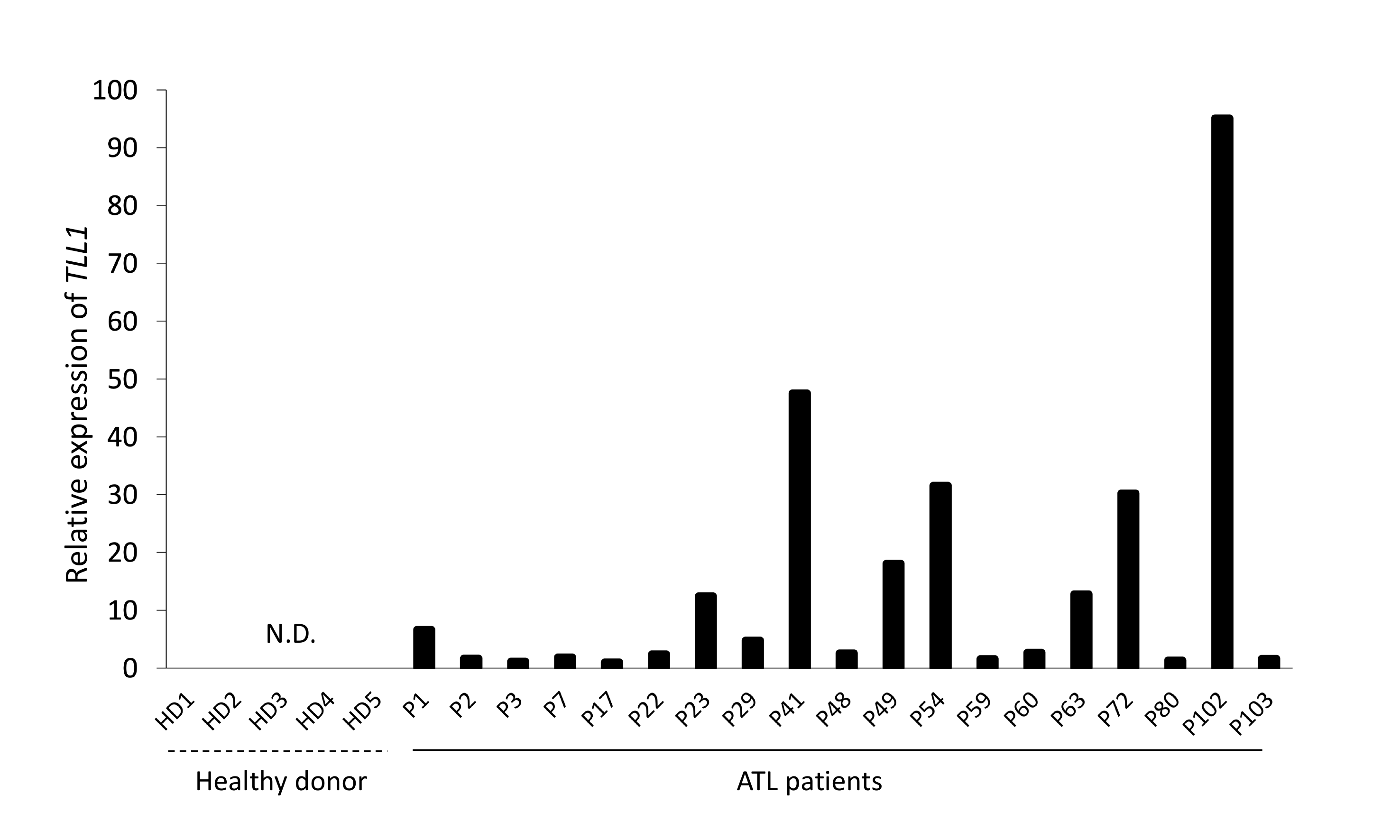}
    \end{minipage}
     
\caption{(a,b) Histogram of ATAC-seq reads around the TLL1 region.
(c) Expression of TLL1 relative to GAPDH (Glyceraldehyde-3-phosphate dehydrogenase) for ATL sample $s$. The relative expression of sample $s$ is defined as $2^{\Delta C_s^0}$, where $\Delta C_s^0$ is computed using the $\Delta \Delta$ Ct method; for details, see \eqref{DDCT}. N.D. stands for no detection of TLL1.}
\label{fig:ATACseq_reads}
\end{figure}

  As shown in Fig. \ref{fig:selected_gene}, the number of key genes was $0$ at $M$=$2000$, $1$ at $M=4000$, $3$ to $4$ between $M=8000$ and $56000$, and then dropped back to $1$ for values larger than $M=64000$. Among these genes, we picked those that stayed in intervals of M $\ge 8000$.  Concretely, TLL1 (Tolloid-like-1) gene appeared from $M=8000$-$16000$, EVC (Ellis-van Creveld) gene and CRMP1 (Collapsin response mediator protein 1) gene appeared from $M=16000$-$48000$, TNRC6A (Trinucleotide Repeat Containing Adaptor 6A) gene appeared from $M=32000$-$48000$, and UST (Uronyl-2-sulfotransferase) gene appeared from $M=64000$. 
  As a reference in Fig.\ \ref{fig:ATACseq_reads}a, we show the locations of the ATAC-seq reads around TLL1 gene.
  Note that the detected genomic region for EVC and CRMP1 is the same because the two genes overlap. Thus, by our selection, the candidates determined as key genes are few.
 
Consistently, EVC has been reported to be overexpressed in ATL and plays an important role in cellular Hedgehog activation \citep{Takahashi2014}. UST was also highly expressed in ATL cases, though the relationship between the function of UST and ATL has not been explicitly clarified \citep{Sasaki2005}.
On the other hand, to the best of our knowledge, TLL1 and TNRC6A have not been studied in this context. 
While TLL1 is known to be necessary for normal septation and positioning of the heart \citep{Clark1999}, a more recent report found that it is associated with the development of hepatocellular carcinoma after the eradication of HCV \citep{Matsuura2017}.
Further, TLL1 is a member of the BMP1/TLD (bone morphogenetic protein1/tolloid)-like proteinase family; BMP1 controls latent TGF-${\beta}$ activation via the cleavage of LTBP1 (latent TGF-$\beta$ binding protein-1) \citep{Ge2006}, and TGF-$\beta$ plays important roles in cancer progression.

Thus, we picked up TLL1 as a promising candidate among genes expressing ATL-specific functions.

\begin{figure}[t]
\begin{center}
  \begin{minipage}[b]{0.9\linewidth}
    \centering
    \begin{flushleft}
        {(a)}
    \end{flushleft}
    \includegraphics[
    trim = 0 0 0 0,
    clip, width=0.7\textwidth]{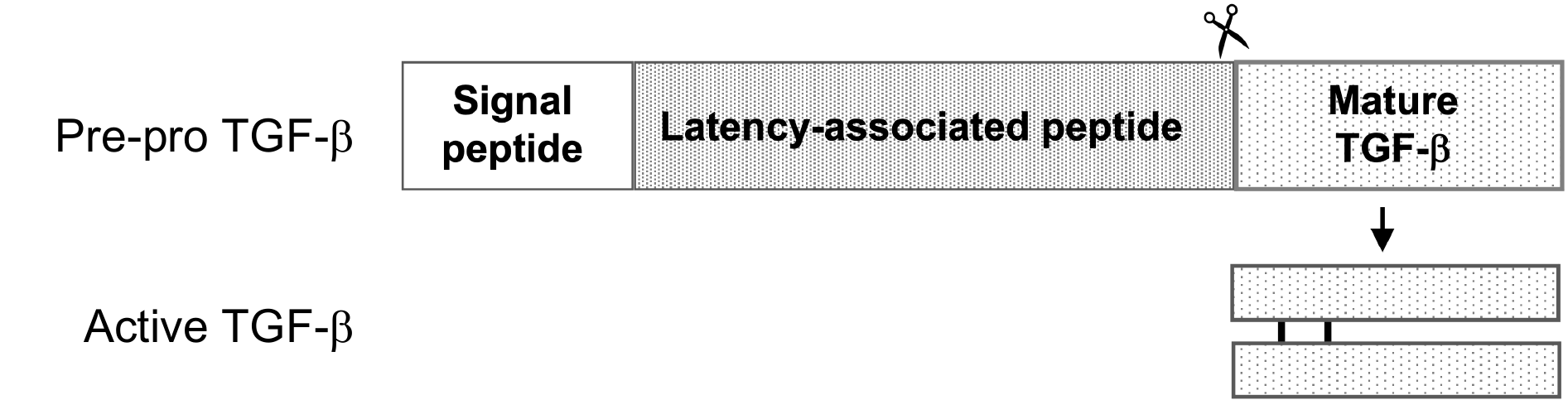}
    \begin{flushleft}
        {(b)}
    \end{flushleft}
    \includegraphics[clip, width=0.6\textwidth]{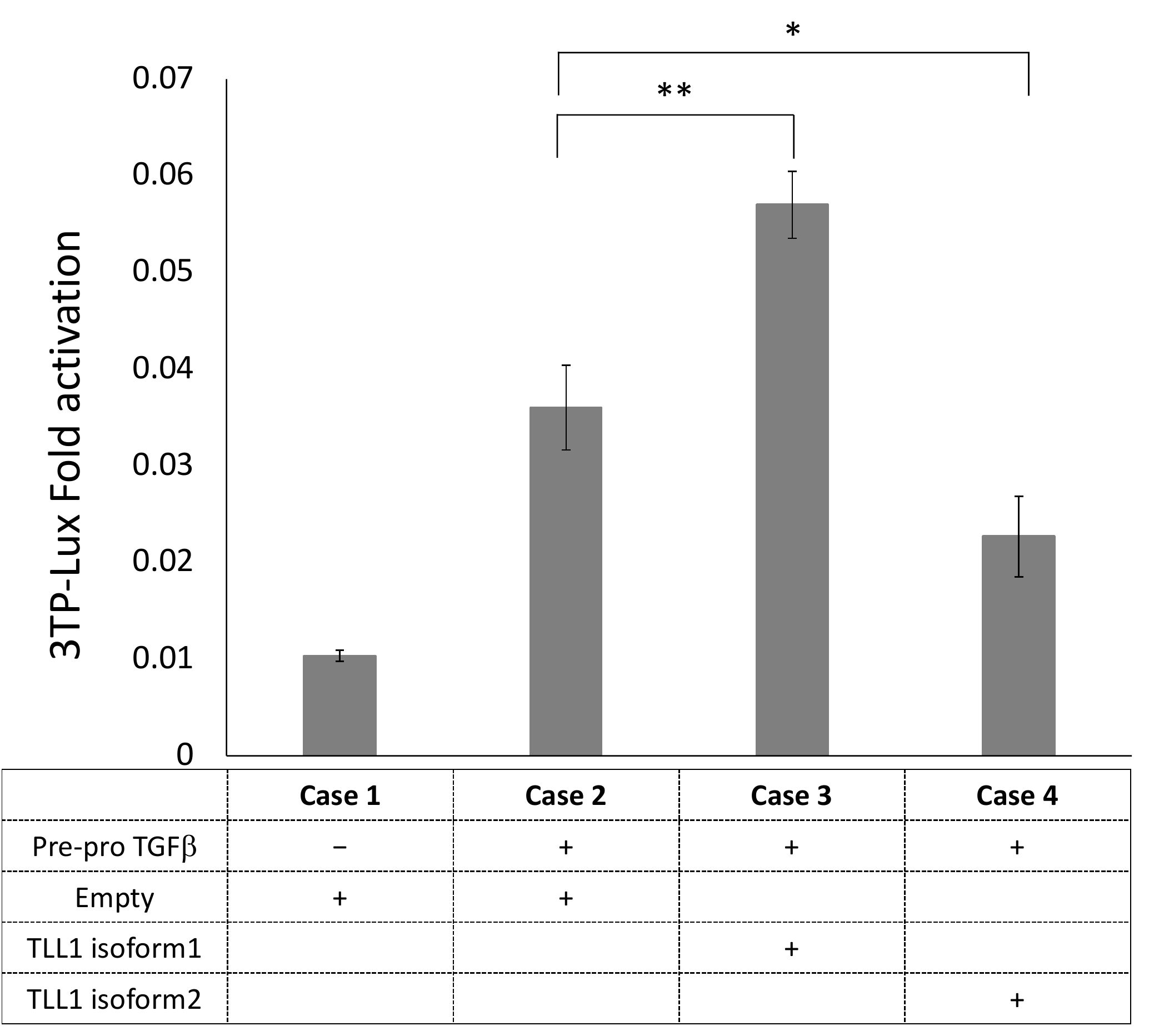}
  \end{minipage}
\end{center}
\caption{
(a) Schematic picture of pre-pro TGF-$\beta$ and mature TGF-$\beta$.
  (b) TGF-$\beta$ activation was measured by 3TP-Lux protein activation, which depended on the TLL1 isoform type. 
  The $p$-value is 0.0028 ($**$) for Case 2 vs. Case 3 and $0.019$ ($*$) for Case 2 vs. Case 4 (t-test). 
  }
\label{fig:TGFbeta}
\end{figure}

\subsection{TLL1 can strongly regulate \texorpdfstring{TGF-$\beta$}{}} \label{Rsection8}

We next considered the gene expression of TLL1 in ATL cells and TLL1's effect on the maturation process of TGF-$\beta$ in a HepG2 line, which is a TGF-$\beta$ responsive cell line\citep{Westerhausen1991}. 

First, as shown in Fig.\ \ref{fig:ATACseq_reads}c, we performed real-time PCR that showed TLL1 is expressed 
in ATL cases but not in the peripheral blood mononuclear cells of healthy donors. 
This result is consistent with the Human Protein Atlas, which shows that TLL1 mRNA is not detected in most adult tissues including immune cells or any hematopoietic lineage.

By analyzing the RNA-seq data, we also confirmed that TLL1 mRNA is not expressed in normal hematopoietic cells, but it was expressed in all examined ATL cases. The same was not true for TNRC6A gene, which, according to RNA-seq data, did not show any systematic expression change between ATL cases and normal hematopoietic cells.

Next, we considered the relationship between TLL1 and the maturation process of TGF-$\beta$. TLL1 has two mRNA isoforms: TLL1 isoform 2 lacks many exons from the 3’ end of TLL1 isoform 1.
Thus, we asked if both isoforms regulate TGF-${\beta}$ in a manner similar to BMP1.

To investigate this possibility, as shown in Fig.\ \ref{fig:TGFbeta}a, we performed a luciferase assay using the pre-mature form of TGF-${\beta}$ co-expressed with either TLL1 isoform 1 or 2 in a HepG2 cell line.
As shown in Fig.\ \ref{fig:TGFbeta}b, we found that compared to the sample without TLL1 (Case 2), TLL1 isoform 1 (Case 3) activates the pre-mature form of TGF-$\beta$ for maturation, but TLL1 isoform 2 (Case 4) represses the maturation.
It should be noted that the difference of luciferase activity between isoform 1 (Case 3) and isoform 2 (Case 4) with pre-mature TGF-$\beta$ approximated the difference between the TLL1-less condition with pre-mature TGF-$\beta$ (Case 2) and without pre-mature TGF-$\beta$ (Case 1). Thus, the results suggest that TLL1 is able to strongly regulate TGF-${\beta}$ depending on the isoform expression ratio.

\section{Discussion}

In this paper, we statistically characterized the anomalous chromatin accessibility and gene expression of HTLV-1-infected cells and healthy CD4$^+$ T cells at the whole genome level. 

Our analysis suggests that compared to healthy CD4$^+$ T cells, ATL cells have the  following properties:
the chromatin accessibility increases near TSSs, a higher frequency of larger sample-to-sample fluctuations at the whole genome level, and a higher frequency of intermediate sample-to-sample fluctuations in gene coding regions. Consistently, histone-related genes were up-regulated. The expression of the genes systematically selected by the chromatin accessibility was found to be distinct from healthy CD4$^+$ T cells but not other hematopoietic cell types that we have studied. 
Further, whereas the immunophenotype determined by the systematically selected open chromatin regions was classified to be near CD4$^+$ T cells for most samples, some samples were classified as myeloid cells.


Based on the above integrative analysis of chromatin accessibility and gene expression, we found that there are chromatin regions which are open in all the ATL cases but closed in all the analyzed samples of the 13 hematopoietic cell types derived from healthy donors. One of the genes overlapping with the chromatin regions that satisfy such conditions is TLL1, which was experimentally shown to have a large potential to regulate TGF-$\beta$.

Contrary to ATL cases, the statistics of the chromatins in HAM cells resembled those of healthy CD4$^+$ T cells, including sample-to-sample fluctuations. This observation implies that there is a certain sample-independent trend in the chromatin structure of HAM cases. 

It should be noted that we were unable to analyze large number of samples due to the difficulty in obtaining samples for given clinical conditions. Thus, some quantities, such as frequencies shown in Fig. \ref{fig:RNA-seq} and Fig. \ref{fig:logFCs}, were not estimated with statistical validity. 
To validate the hypothesis about the uniqueness of ATL cells across different scales from chromatin and transcription to immunophenotypes, more samples should be used in future studies.

Our finding about ATL samples might motivate us to consider a rather general relationship between increased chromatin accessibility and the onset of leukemia. Indeed, in a previous study of Acute Myeloid Leukemia (AML), it was reported that mutations in cohesin genes increase chromatin accessibility, which controls the activity of transcription factors leading  to leukemogenesis \citep{Mazumdar2015}. It was also reported that HMGN1 amplification is also associated with increased chromatin accessibility; It confers a transcriptional and chromatin phenotype associated with stem cells and leukemia \citep{ Cabal-Hierro2020}. It remains for future studies to check how the relationship between the increased chromatin accessibility and the onset of leukemia can be generalized beyond the above cases such as ATL and AML.

Let us discuss our additional findings in the following, though they are preliminary results. First, to compare ATL cases with another type of leukemia in terms of immunophenotypes, we analyzed the chromatin structure of cutaneous T-cell lymphoma (CTCL) using the ATAC-seq data of CTCL (GSE85853) \citep{Tanaka2020}, which is reported to have a clinical and histopathological phenotype similar to that of ATL \citep{Qu2017}.  
As shown in Table \ref{tab3}, we found that the chromatin structures in some CTCL cases are closer to myeloid cells rather than CD4$^+$ T cells, though CTCL is conventionally classified as T-cell leukemia. Especially, in the cases of Patient 11 and 60, both of whom are romidepsin responders, the chromatin structure changed from myeloid cell-like to CD4${^+}$ T-like after drug treatment on the 0th day.
Although the molecular mechanisms underlying this process are not understood, these results suggest that the change in immunophenotype reflects a molecular response in treatment. This observation could shed light on finding better therapeutic targets and prediction of drug response. 

\begin{table}[hbtp]
\caption{Clustering results of ATAC-seq from 9 CTCL cases in terms of immunophenotype as a function of time using the method in \citep{Tanaka2020}. 
The Histone deacetylase inhibitor (HDACi) was romidepsin. $+$ and $-$, positive response and negative response, 
respectively \citep{Qu2017}.}
\label{tab3}
 \begin{tabular}{|l|c|c|c|c|c|}
\hline
SRR  & first & second  & third & Patient tag  & HDACi\\
Number &  &   &  & \& Time tag  & responder\\
\hline   
4044872 & Ery & Mono & CLP & Patient-11 on 0-th day & \\
4044873 & CD4$^+$ T & CD8$^+$ T  & Ery & on 0-th day &\\
4044874 & CD4$^+$ T  & CD8$^+$ T  & NK &  on 7-th day & $+$\\
4044875 & CD4$^+$ T & CD8$^+$ T & NK &  on 7-th day &\\
4044876 & CD4$^+$ T  & CD8$^+$ T & B &  on 14-th day  &\\
4044877 & Ery & CD4$^+$ T & Mono &  on 14-th day  &\\
\hline
4044878 & Ery & Mono & CLP & Patient-20 on 7-th day & $-$\\
\hline
4044879 & Ery & Mono & CLP & Patient-39 on 0-th day  & \\
4044880 & Ery & Mono & CLP  &  on 0-th day & $-$\\
4044881 & Ery & Mono & CLP  &  on 7-th day &\\
4044882 & Ery & Mono & CLP  &  on 7-th day &\\
\hline
4044885 & CD4$^+$ T & CD8$^+$ T & B & Patient-59 on 7-th day & $+$\\
4044886 & CD4$^+$ T & CD8$^+$ T & B &  on 7-th day &\\
\hline
4044887 & Ery & Mono & CD4$^+$ T & Patient-60 on 0-th day & $+$\\
4044888 & CD4$^+$ T & B &  CD8$^+$ T &  on 7-th day  &\\
\hline
4044889 & Ery & Mono & CLP & Patient-61 on 0-th day  & $+$\\
\hline
4044890 & CD4$^+$ T & CD8$^+$ T & B & Patient-62 on 0-th day & $+$\\
4044891 & CD4$^+$ T & Ery & CD8$^+$ T  &  on 0-th day &\\
\hline
4044892 & CD4$^+$ T & CD8$^+$ T & NK & Patient-1366 on 0-th day & $+$\\
\hline
\end{tabular}
\end{table}

\begin{figure}[t]
\begin{center}
 \begin{minipage}[b]{0.4\linewidth}
     \centering
     \begin{flushleft}
     {(a) ETS1}
     \end{flushleft}
    \includegraphics[clip, width=\linewidth]{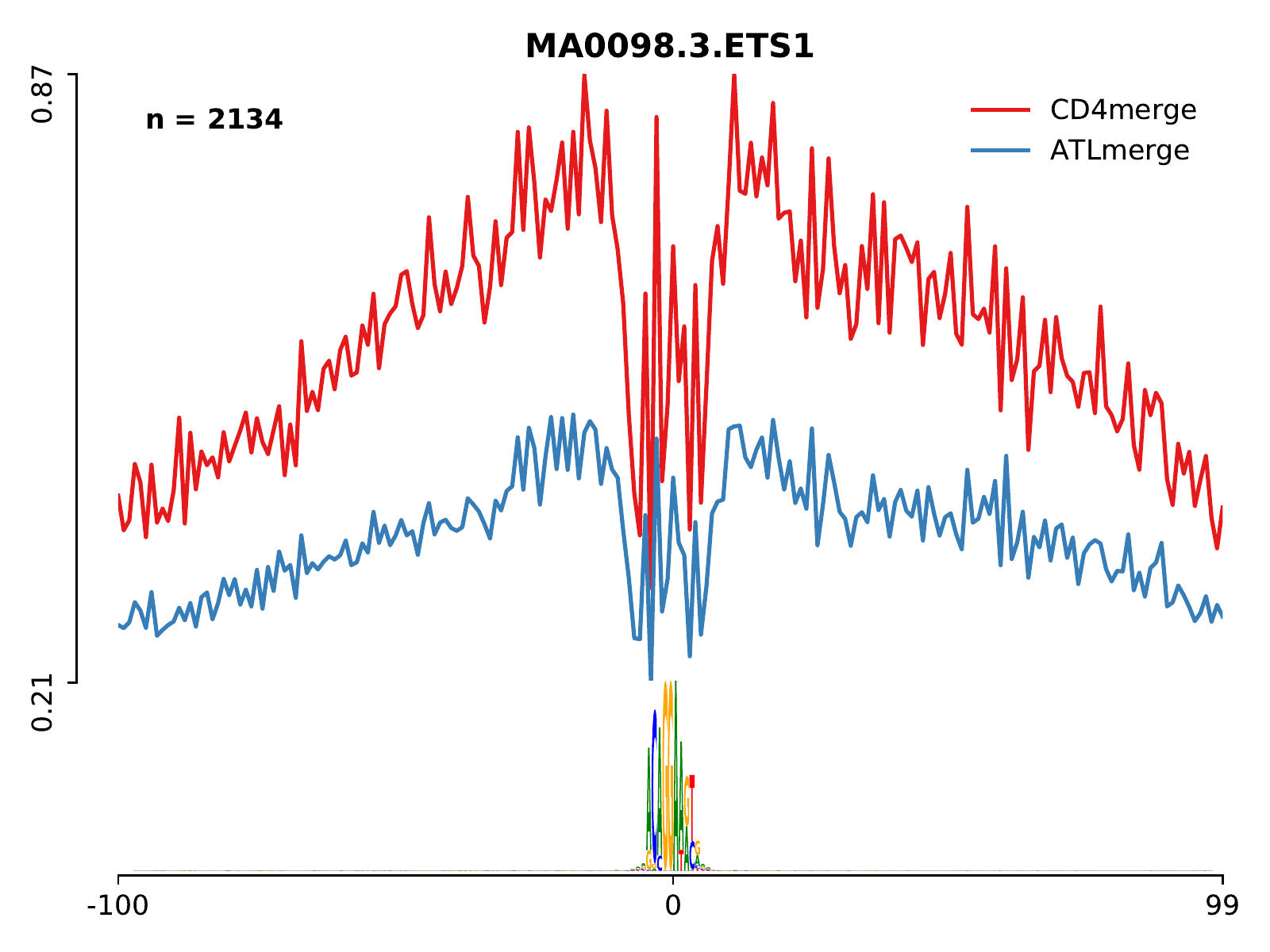}
    \end{minipage}
    \hfill
    \begin{minipage}[b]{0.4\linewidth}
      \centering
      \begin{flushleft}
      {(b) IRF2}
      \end{flushleft}
    \includegraphics[clip, width=\linewidth]{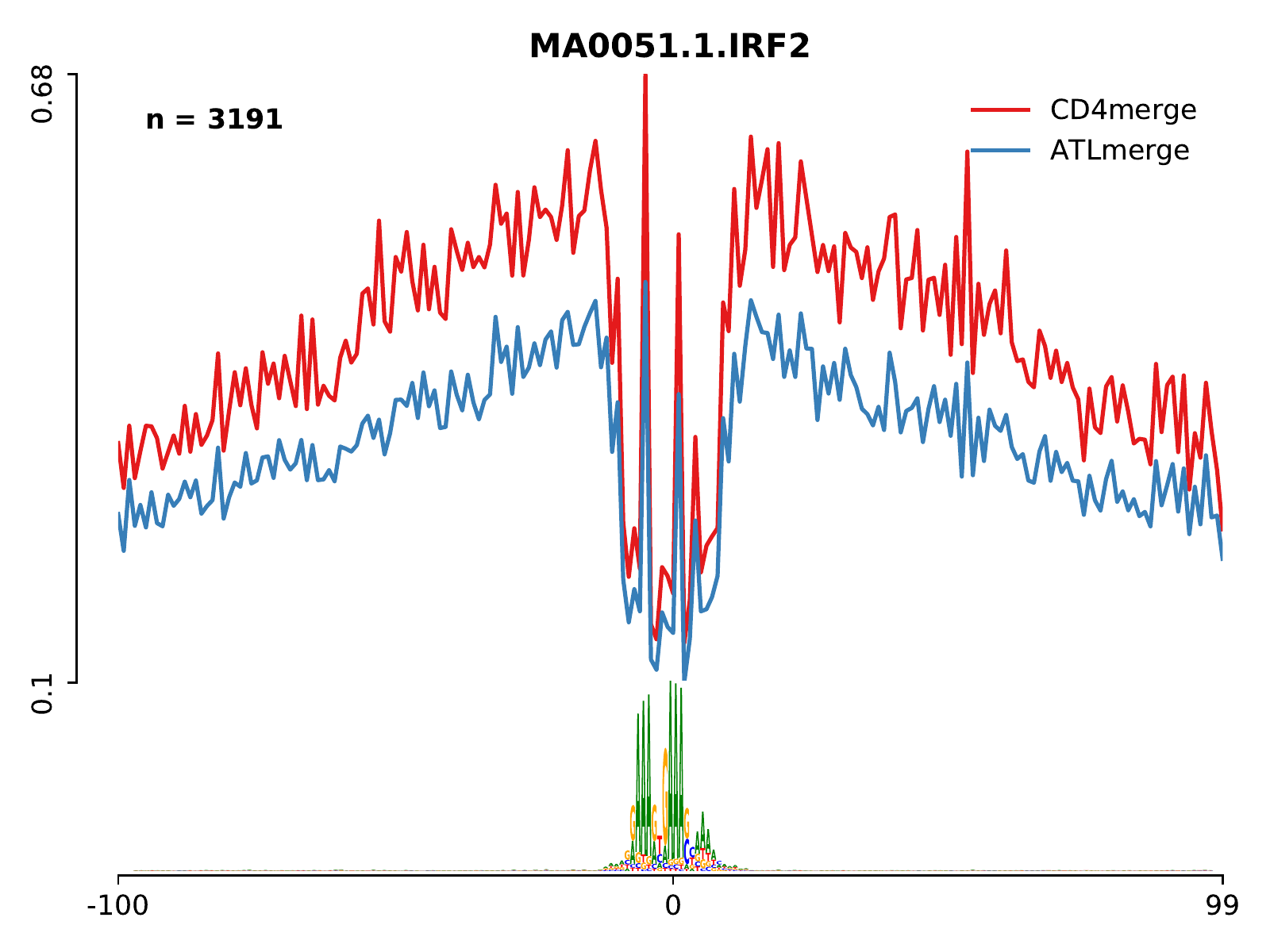}  
    \end{minipage}    

\vspace{15pt}
    
    \begin{minipage}[b]{0.4\linewidth}
     \centering
     \begin{flushleft}
{(c) RUNX2}
\end{flushleft}
    \includegraphics[clip, width=\linewidth]{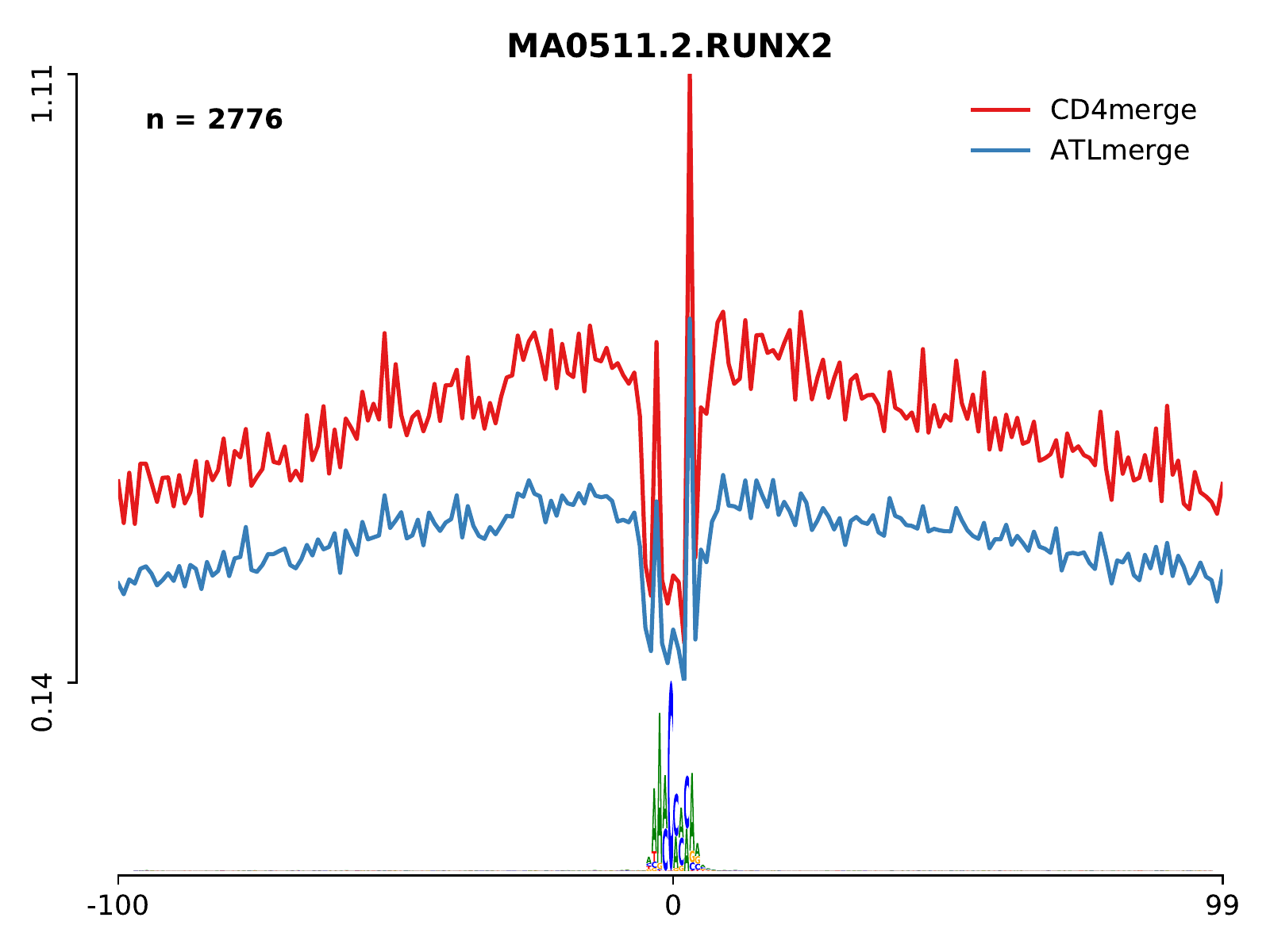}
    \end{minipage} 
    \hfill
    \begin{minipage}[b]{0.4\linewidth}
     \centering
     \begin{flushleft}
     {(d) NRF1}
     \end{flushleft}
 \includegraphics[clip, width=\linewidth]{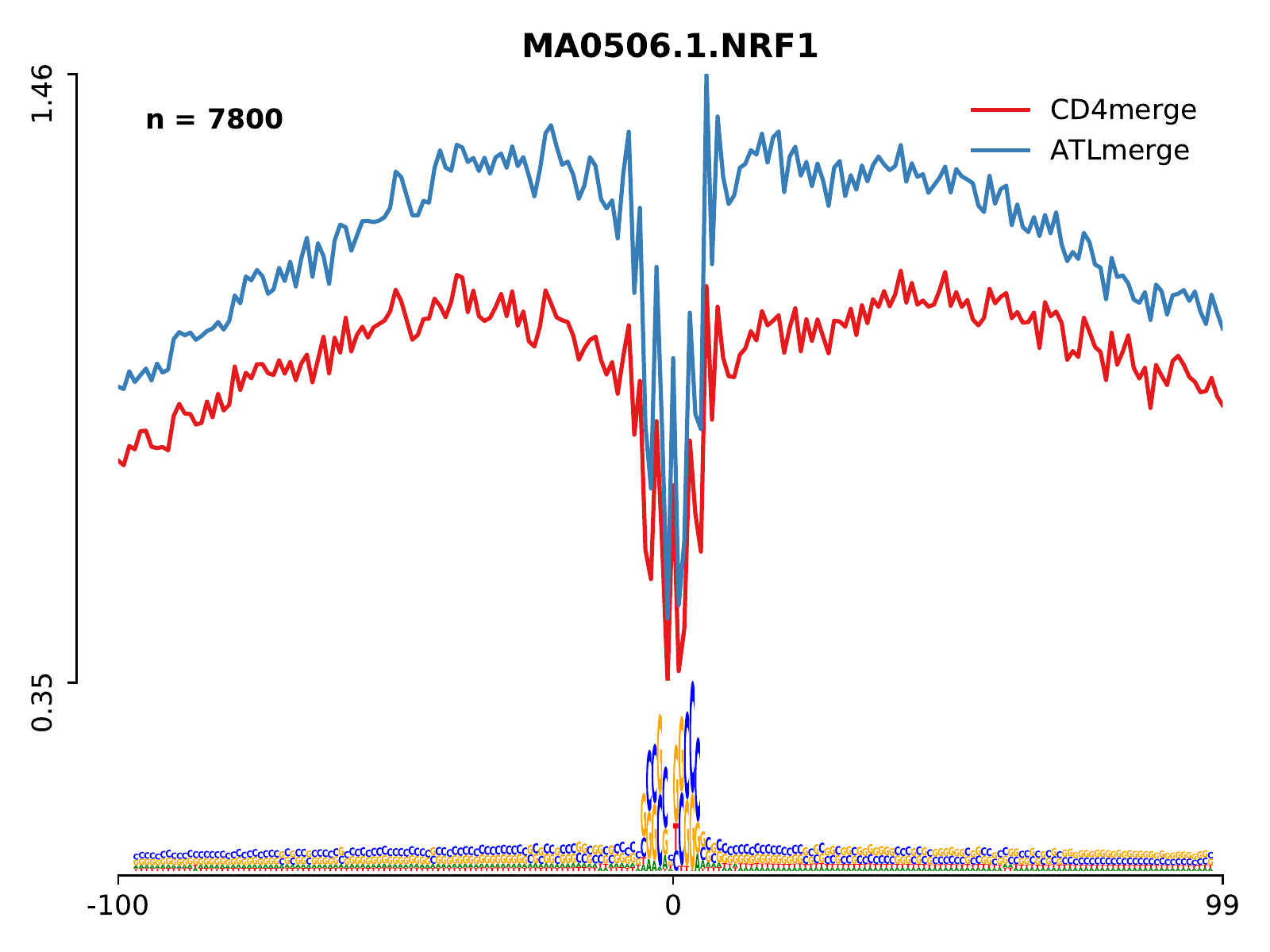}
    \end{minipage}

\vspace{15pt}
    
    \begin{minipage}[b]{0.4\linewidth}
     \centering
     \begin{flushleft}
     {(e) KLF4}
     \end{flushleft}
   \includegraphics[clip, width=\linewidth]{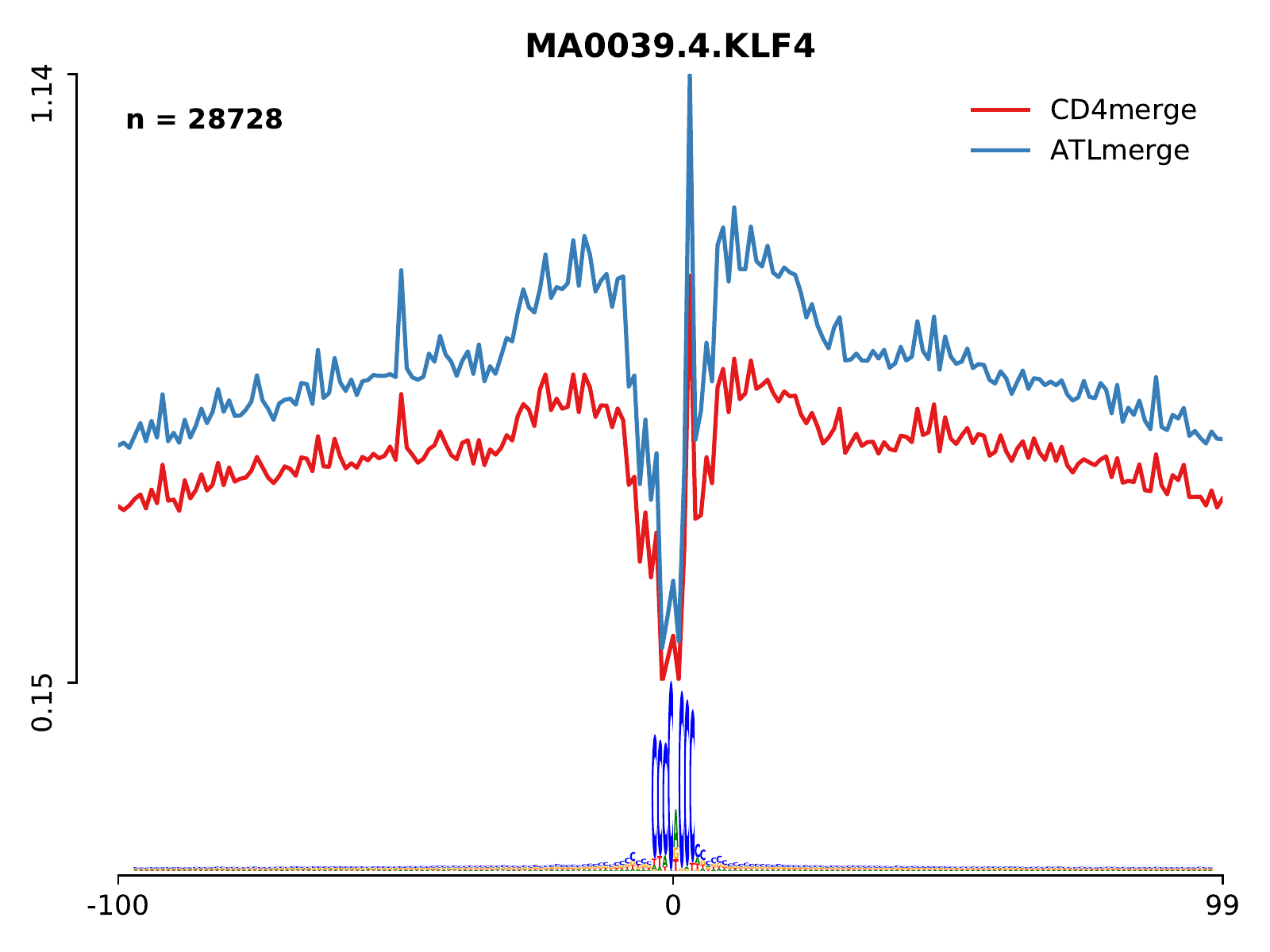}
    \end{minipage} 
    \hfill
    \begin{minipage}[b]{0.4\linewidth}
     \centering
     \begin{flushleft}
    {(f) KLF9}
    \end{flushleft}
     \includegraphics[clip, width=\linewidth]{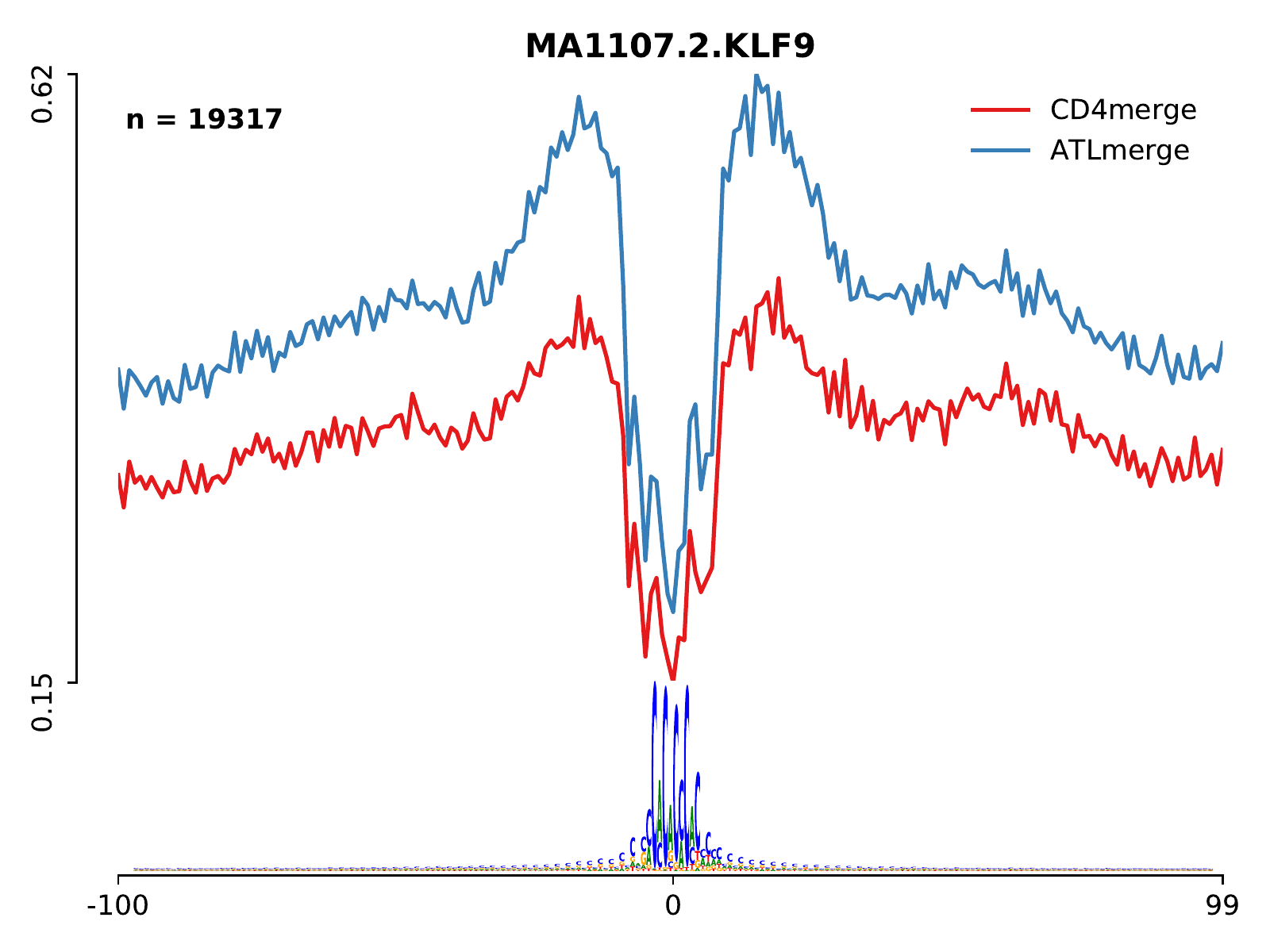}
    \end{minipage}
\end{center}
 \caption{A footprint analysis of transcription factors. (a)-(f) Distances from motifs (horizontal axis) vs. the averaged number of reads over all parts of a given motif (vertical axis) outputted from HINT-ATAC \citep{Li2019}.}
\label{fig:footprints}
\end{figure}

Second, looking at the ATAC-seq data further, a footprint analysis for the identification of differential motifs revealed that ETS1, IRF2, and RUNX2 had deeper footprints and higher DNA accessibility at the flanking locations of their motifs in healthy CD4${^+}$ T cells, while NRF1, KLF4, and KLF9 had deeper footprints in ATL cells (Fig. \ref{fig:footprints}) \citep{Li2019}. These observations suggest that transcription factors such as NRF1, KLF4, and KLF9 play an important role in ATL. 

Third, to understand TLL1's function, we conducted experimental studies on MT-2, an ATL cell line that does not express TLL1. We prepared three samples of MT-2: one transduced with an empty vector, another with TLL1 isoform 1, and the third with TLL1 isoform 2. We analyzed the RNA-seq data from the three samples 
by computing the gene expression ratio between them.

\begin{table}[hbtp]
\caption{Relative expression (RE) of genes in TLL1-transduced MT-2 cells over MT-2 cells transduced with an empty vector. Type 1(2) corresponds to MT-2 cells transduced with TLL1-isoform 1 (isoform 2). Genes were picked up if the RE of type 1 was larger than $11$ and the RE of type 2 was smaller than $3$ or if the RE of type 1 was smaller than $1$ and the RE of type 2 was larger than $5$.}
\label{TLL1var1}
 \begin{tabular}{|c|c|c|}
\hline
Gene symbol  & RE of type 1  & RE of type 2 \\
\hline   
CCL3 & 14.53 & 2.94 \\
\hline
CCR6 & 11.70 & 2.85 \\
\hline
MIR155 & 11.30 & 2.44 \\
\hline
POSTN & 11.06 & 2.49  \\
\hline
HBG2 & 0.058 & 15.00 \\
\hline
HBB & 0.15 & 12.29 \\
\hline
HBA2 & 0.14 & 5.79 \\
\hline
HBA1 & 0.09 & 5.04 \\
\hline
\end{tabular}
\end{table}

Among the $22963$ genes, 
we picked up the genes which have nonzero reads count in MT-2 transduced with an empty vector. 
A part of such genes showed significantly altered expressions depending on the type of isoform. As shown in Table \ref{TLL1var1}, four genes were significantly up-regulated when TLL1-isoform 1 was transduced: CCR6 is related to the regulation of Treg migration \citep{Yamazaki2008}, microRNA-155 (MIR155) modulates Treg cell differentiation and its expression is up-regulated in HTLV-1 transformed cells \citep{Yao2012, Pichler2008, Watanabe2017}, the chemokine CCL3 regulates myeloid differentiation \citep{Staversky2018}, and POSTN has been reported to be involved in TGF-${\beta}$ activation \citep{Sidhu2010}.
As for the four genes significantly up-regulated when TLL2-isoform2 was transduced, all were globin genes.
These findings reiterate the dependence of TLL's function on its isoforms in ATL cells. 

As an additional note, Fig.\ \ref{fig:ATACseq_reads}b suggests that the chromatin regions around the TLL1 loci tend to be open also for the HAM cases. This observation indicates that open chromatin regions around TLL1 are not the only cause of leukemia onset. Rather, it suggests that open chromatin regions are potentially caused by the infection itself and related to the latent period or expansion of the virus. 


This study is the first comprehensive analysis of open chromatin structures in ATL samples. The findings will deepen understanding of the ATL pathogenesis.

\section{Methods and Materials}
\begin{figure}[t]

\includegraphics[keepaspectratio, width=0.9\linewidth]{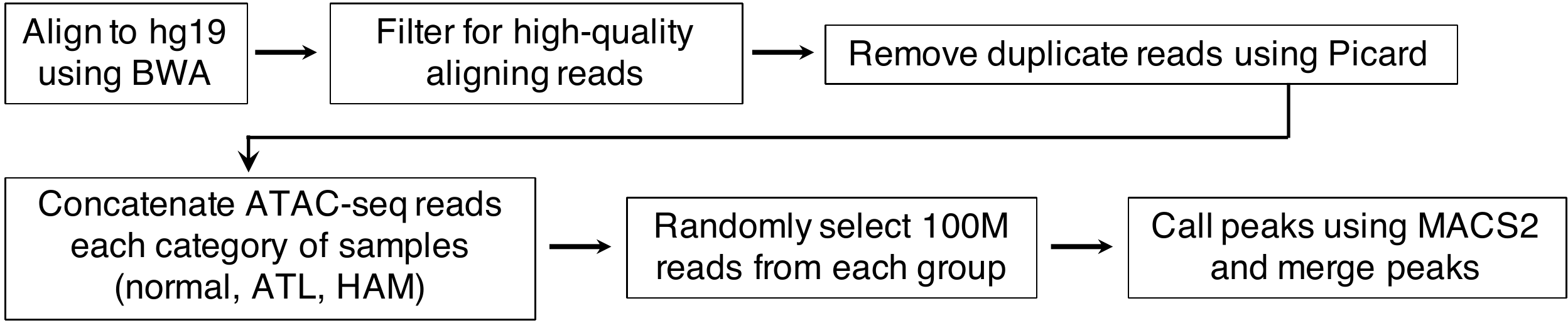}
\caption{Pipeline of the data processing for ATAC-seq.}\label{fig:ATAC_pipeline}
\end{figure}

\subsection{Sequencing sample preparation} \label{MMsection1}
Peripheral blood mononuclear cells from ATL patients, HAM patients, and HTLV-1 carriers were thawed and washed with PBS containing 0.1\% BSA. To discriminate dead cells, we used a LIVE/DEAD Fixable Dead Cell Stain Kit (Invitrogen). For cell surface staining, cells were stained with APC anti-human CD4 (clone: RPA-T4) (BioLegend) and anti-SynCAM (TSLC1/CADM1) mAb-FITC (MBL) antibodies for 30 minutes at 4 ℃ followed by washing with PBS. HTLV-1-infected cells (CAMD1${^+}$ and CD4${^+}$) were purified by sorting with a FACS Aria (Beckman Coulter) to reach 98–99\% purity. Data were analyzed using FlowJo software (Treestar). Soon after the sorting, 10000-50000 HTLV-1-infected cells were centrifuged and used for ATAC-seq. Total RNA was isolated from the remaining cells using an RNeasy Mini Kit (Qiagen). Library preparation and high-throughput sequencing were performed at Macrogen Inc. (Seoul, Korea). The diagnostic criteria and classification of the clinical subtypes of ATL were performed as previously described \citep{Shimoyama1991}. 77 ATAC-seq datasets from 13 human primary blood cell types were obtained from the Gene Expression Omnibus (GEO) with accession number GSE74912 \citep{Corces2016}, and RNA-seq datasets of CD4${^+}$ T and CD8${^+}$ T cells from healthy donors were obtained from GSE74246 \citep{Corces2016}. 
The RNA-seq data for ATL samples can be downloaded from DBBJ (DNA Data Bank of Japan) with accession number DRR250721 for ATL8, DRR250717 for ATL10, DRR250718 for ATL21, and DRR250719 for ATL24. The ATAC-seq data for ATL samples can be downloaded from DBBJ with accession number DRR250714 for ATL8, DRR250710 for ATL10, DRR250711 FOr ATL21, and DRR250712 for ATL 24. 

\subsection{Pre-processing of ATAC-seq} \label{MMsection2}
High-throughput sequencing provides a set of reads as output. 
ATAC-seq reads were aligned using BWA version 0.7.16a with default parameters. SAMtools was used to convert SAM files into compressed BAM files and sort the BAM files by chromosome coordinates. PICARD software (v1.119) (\url{http://broadinstitute.github.io/picard/}) was used to remove PCR duplicates using the \texttt{MarkDuplicates} options. Reads with mapping quality scores less than 30 were removed from the BAM files. For peak calling, MACS2 (v2.1.2) software was used with option \texttt{{-}{-}nomodel {-}{-}nolambda {-}{-}keep-dup all -p 0.01}. ATAC-seq tracks were visualized using Integrative Genomics Viewer (IGV), and footprinting analysis was performed using HINT-ATAC \citep{Li2019}. Note that  the paired-end output of the sequence was used to reconstruct the fragments, where paired two reads correspond to both ends of a fragment.

\subsection{Pre-processing of RNA-seq}
RNA-seq data were aligned to human reference genome hg19 using STAR 2.6.0c with the \texttt{--quantMode} GeneCounts function \citep{Dobin2013}. RNA-seq data analysis was performed using edgeR, where the reads counts of the RNA-seq data were normalized using TMM normalization \citep{cite-key} to be converted into pseudo reads counts. Let $n_i^0(s)$ be the reads count of the RNA-seq data for each gene $i$ for a given cell sample $s$ and $N(s):=\sum_i n_i^0(s)$ be the total reads count over all genes.
Using the TMM normalization with $n_i^0(s)$ for gene $i$ and sample $s$ as the input data,
one can obtain the normalization factor $r(s)$ for a given sample $s$ using the command \texttt{calcNormFactors} of edgeR.
After acquiring $r(s)$,
the pseudo reads count $n_i(s)$ is calculated using the command \texttt{estimateCommonDisp} of edgeR, which we used as the starting point of the RNA-seq data analysis in the main text.

An additional analysis was done to evaluate robustness. We computed the geometric mean of $N(s)r(s)$ as $N_0= \left( \prod_{s \in \mathbb{S}} r(s)N(s) \right)^{1/|\mathbb{S}|}$, where $\mathbb{S}$ is the set of samples and
$|\mathbb{S}|$ is the number of elements of $\mathbb{S}$. 
We then checked whether the pseudo reads count $n_i(s)$ is close to
a normalized reads count $n_i'(s):=n_i^0(s)\frac{N_0}{r(s)N(s)}$ for sample $s$. 
We found that
the maximum deviation between $n_i'(s)$ and $n_i(s)$ is smaller than $5$ over all genes in the case of CD4$^+$ T vs. Mono in Fig.\ \ref{fig:logFCs}a. 
In this case, the effects from the differences between $n_i'(s)$ and $n_i(s)$ are quite small except for quantities related to genes with almost zero reads. Therefore, even if we use $n_i'(s)$ as the starting point of the analysis 
instead of the pseudo reads count $n_i(s)$, qualitatively the same conclusion as that with $n_i(s)$ is expected to be obtained.

A PCA was done using the covariance matrix of $\log_{10} (n_i(s)+1)$, where the first and second principal components were calculated using the  \texttt{prcomp} command with option \texttt{scale=FALSE} in R.

\subsection{Cell lines and clinical samples}\label{MMsection3}
All ATL cell lines were cultured in RPMI 1640 medium supplemented with 10\% FBS and antibiotics. HepG2 was cultured in DMEM. To construct MT-2 cells stably expressing TLL1 isoform 1 or 2, the coding sequence of each isoform was transduced using a lentivirus vector constructed as described in subsection \ref{MMsection4}.

\subsection{Real-time PCR}\label{MMsection7}
cDNA products were analyzed by real-time PCR using PowerUp SYBR Green Master Mix and StepOnePlus Real-Time PCR System (Applied Biosystems) according to the manufacturer's instructions. Primer sequences for the GAPDH gene have been described previously \citep{Ponchel2003}, and primer sequences for the TLL1 gene were 5- TTGTTTTCTACGGGGAGCTATGG-3 and 5- ATATCGCCCCAAAATACAGCG-3. The relative quantification was calculated according to the method described in Applied Biosystems ABI prism 7700 SDS User Bulletin \#2.
Note that the ATL sample used for this experiment is not listed \rd{in} Table \ref{tab2}.

\subsection{Lentiviral vector construction and transfection of recombinant lentivirus}\label{MMsection4}
The coding region of TLL1 isoform 2 was synthesized using gBlocks Gene Fragment (Integrated DNA Technolofies), which was used as the template for synthesizing TLL1 isoform 1 by PCR amplification. TLL1 isoform 1 and 2 fragments were subcloned into pCS2-EF-MCS (gift from H. Miyoshi, RIKEN Bioresource Center). An empty vector that expresses only hrGFP was used as the control for the lentiviral transduction. 293T cells at 80\% confluence in a 10-cm dish were co-transfected with 10 ${\mu}$g lentivirus vector, 10 ${\mu}$g psPAX2, 5 ${\mu}$g pMD2.G, and PEI (Polyethylenimine). 48 hours after the transfection, supernatant containing the virus was collected and concentrated by ultracentrifugation. MT-2 cells were transfected with the lentivirus, and two weeks after the transduction, GFP-positive cells were purified by sorting with FACS Canto. RNA was isolated using the Qiagen RNeasy Mini Kit and then used for the RNA-seq analysis.

\subsection{Luciferase assay}\label{MMsection5}
The coding region of human TGF-${\beta}$, whose length is 1173 bp, was generated by PCR amplification and subcloned into a pFUSE-hIgG1-Fc2 vector. HepG2 cells were plated on 12-well plates at ${1\times10^5}$ cells per well. After 24 hours, the cells were transfected with 50 ng/well of luciferase reporter plasmid (p3TP-Lux) \citep{Zhao2011}, 5 ng/well of Renilla luciferase control vector (phRL-TK) together with 35 ng/well of TLL1 expression plasmid, and 35 ng/well of TGF-${\beta}$-expressing plasmid or empty vector. Plasmids were transfected using TransIT-LT1 (Mirus) according to the manufacturer’s instructions. After 48 hours, the cells were collected, and luciferase activities were measured using the Dual-Luciferase Reporter Assay Kit (Promega). Relative luciferase activity was calculated as the ratio of firefly to Renilla luciferase activity. Three independent experiments, each with triplicate transfections, were performed, and typical results are shown.

\subsection{Explicit definitions of the computed quantities}

We explicitly define quantities discussed in the main text. First, we assume that the set of reads from DNA of an ATAC-seq sample and 
the reads count for each gene from an RNA-seq sample are given. Also, a set of fragments for the ATAC-seq sample is also given by using a pair of reads; both ends of a fragment correspond to a pair of two reads. The positions of TSSs and coding regions of all genes were obtained from the human genome (hg19) 
as a set of intervals on the genome. 
Therefore, a read from the ATAC-seq data is interval $[x_1,x_2]$ on the genome, which corresponds to a region including an edge of a fragment. A fragment has length $\ell$ and location $x$ as the mid-point of the two edges on the genome \citep{Tanaka2020}. The reads from the RNA-seq data provide the read count $n_i(s)$ for each gene $i$, where $s\in \mathbb{S}$ is the sample index. We denote by $\mathbb{S}_\nu$
the set of all analyzed samples with type $\nu$.

\subsubsection{The normalized number of reads in Fig.\ \ref{fig:around_TSS}a: }
Let us consider the number $\rho_s(z)$ of ATAC-seq reads from a sample $s \in \mathbb{S}$ located on position $z$ from the nearest TSS. 
Then, we take the sample average among type $\nu$ as
\begin{align}
  \overline{\rho}_{\nu}(z):=\dfrac{1}{|\mathbb{S}_\nu|}\sum_{s\in \mathbb{S}_\nu}
  \rho_{s}(z),
\end{align}
where $\nu\in\{\mathrm{CD4^+T},\mathrm{HAM},\mathrm{ATL}\}$.
In Fig.\ \ref{fig:around_TSS}a, we plot the normalized quantity $\widetilde{\rho}_{\nu}(z)$ obtained after dividing $\overline{\rho}_{\nu}(z)$ by the value at the TSS ($z=0$) such that
\begin{align}\label{eq:1d_TSS_quantity}
  \widetilde{\rho}_{\nu}(z):=\dfrac{\overline{\rho}_{\nu}(z)}{\overline{\rho}_{\nu}(0)}.
\end{align}

\subsubsection{The averaged number of fragments in Fig.\ \ref{fig:around_TSS}b: }
Let us consider the number of fragments $\phi_s(z, \ell)$ from sample $s\in\mathbb{S}$ satisfying the following two conditions:
(1) their centers are located at $z$, and (2) they have length $\ell$.
$\overline{\phi}_\nu(z, \ell)$ describes the sample average among type $\nu$ as
\begin{align}
  \overline{\phi}_\nu(z, \ell) :=
  \frac{1}{|\mathbb{S}_\nu|} \sum_{s \in \mathbb{S}_\nu} \phi_s(z, \ell),
\end{align}
where $\nu\in\{\mathrm{CD4^+T},\mathrm{HAM},\mathrm{ATL}\}$.
In Fig.\ \ref{fig:around_TSS}b, we plot the histogram $F^{\Delta, \xi}_{\nu}(z, \ell)$ for bin width $\Delta$ and $\xi$ for $z$ and $\ell$, respectively, as
\begin{align}\label{eq:2d_TSS_quantity}
  F^{\Delta,\xi}_{\nu}(z,\ell):= \sum_{z - \Delta/2 \le z' < z + \Delta/2} 
  \sum_{\ell \le \ell' < \ell + \xi}
  \overline{\phi}_{\nu}(z',\ell').
\end{align}

\subsubsection{The reference set of peaks:}\label{subsub:PeakData}
To analyze open chromatin regions, we used MACS2 with the input of reads from the ATAC-seq data. Concretely, we used MACS2 with the option \texttt{{-}{-}nomodel {-}{-}nolambda {-}{-}keep-dup all -p $0.01$}, which corresponds to $p_G=10^{-2}$ \citep{Tanaka2020}.
This algorithm outputs the collection of peaks $\hat{g}_s$ for a given sample $s$ as candidates of open chromatin regions, which can be described as
\begin{align} 
\hat{g}_s := ((\gamma_k, \alpha_k,\beta_k),p_k)_{k \ge 1},
\end{align} 
where $\gamma_k$ is the chromosome number, $\alpha_k$ is the starting point, and $\beta_k$ is the ending point in terms of genome position 
with $p_k$ as the $p$-value of the $k$-th peak. 
As in \citep{Tanaka2020}, 
$p_k \le p_{k'} $ for $k<k'$. 
In particular, we consider the set of the top $M$ peaks and denote it by
\begin{align}
    \hat{g}_{s}^M := \begin{cases}
    ((\gamma_k, \alpha_k,\beta_k),p_k)_{k=1}^{M} & (\text{if $|\hat{g}_s| \ge M$}), \\
    \hat{g}_s & (\text{otherwise}).
    \end{cases}
\end{align}

Next, we concatenate the data of all reads from all ATAC-seq samples with cell type $\nu\in\{\mathrm{CD4^+T},\mathrm{HAM},\mathrm{ATL}\}$. 
Then, we randomly extract 100 million reads from the concatenated data for type $\nu$ as the input for the MACS2 algorithm to obtain the collection of peaks, 
\begin{align} 
g_{\nu} := ((\gamma_k, \alpha_k,\beta_k),p_k)_{k \ge 1}.
\end{align}
Using a coalescing process of 
$g_{ATL}, g_{HAM}, g_{CD4^+T}$, we construct a new reference set of peaks $g_0$ as follows.
Operationally, the coalescing of two peaks is done as follows. 
If two peaks $(\gamma, \alpha,\beta)$ and $(\gamma', \alpha',\beta')$ for $\gamma=\gamma'$ satisfy $\alpha'\le \alpha \le \beta'$, the two peaks become one peak as $(\gamma, \alpha', \max\{\beta,\beta'\})$. 
This operation is repeated for a newly obtained set of peaks 
until no more coalescing processes occur. 

\subsubsection{The length of overlapped peaks in Fig.\ \ref{fig:averages}, \ref{fig:variances}::}
To quantify the similarity between two collection of peaks, $g_0$ and $\hat{g}_s^M$, first, we fix a set of peaks $\mathbb{L}$ as any of
(1) the set $\mathbb{G}$ of all peaks in $g_0$ overlapping gene coding regions,
(2) the set $\mathbb{G}^c$ of all peaks in $g_0$ corresponding to non-coding regions, and (3) the union $\mathbb{G} \cup \mathbb{G}^c$ of two sets $\mathbb{G}, \mathbb{G}^c$. Note that for a given peak, its center was calculated by the command of \texttt{annotatePeaks.pl} in the HOMER algorithm and used to judge whether the peak joins $\mathbb{G}$ or $\mathbb{G}^c$ \cite{Heinz2015}.

Then, we focus on the length of the overlapped peaks $O_k^{\mathbb{L}} (g_0, \hat{g}_s^M)$, which is the number of base pairs in a peak of $\hat{g}_s^M$ inside the $k$-th peak $(\alpha_k, \beta_k)$ of $\mathbb{L}\subset g_0$. Then, we compute the average and variance of $O_k^{\mathbb{L}} (g_0, \hat{g}_s^M)$ as follows:
\begin{eqnarray}
  &\overline{O}_k(\mathbb{L}, \mathbb{S}):=\dfrac{1}{|\mathbb{S}|} \sum_{s\in \mathbb{S}} O_k^{\mathbb{L}} (g_0, \hat{g}_s^M),\\
  &V_k(\mathbb{L}, \mathbb{S}) := \dfrac{1}{|\mathbb{S}|}\sum_{s\in \mathbb{S}}\big(O_k^{\mathbb{L}}(g_0,\hat{g}_s^M)-\overline{O}_k(\mathbb{S}, \mathbb{L})\big)^2.
\end{eqnarray}
We set $M=64000$ as the provisionally optimal number for immunophenotype classification \citep{Tanaka2020}.

The following functions describe the frequency of the average and variance of $O_k^{\mathbb{L}} (g_0, \hat{g}_s^M)$:

\begin{align}
\rho^{(1)}_{\mathbb{L},\mathbb{S}}(O) := \sum_{k\in\mathbb{L}}\delta(O,\overline{O}_k(\mathbb{L}, \mathbb{S})),\\
\rho^{(2)}_{\mathbb{L},\mathbb{S}}(V) := \sum_{k\in\mathbb{L}}\delta(V,V_k(\mathbb{L}, \mathbb{S})),
\end{align}  
where $\delta(a,b)=1$ for $a=b$, otherwise $0$. 

Lastly, in Fig.\ \ref{fig:averages} and Fig.\ \ref{fig:variances}, 
the histograms are defined as
\begin{align} 
F^{(1)}_{\mathbb{L},\mathbb{S}}(O;\Delta):= \sum_{O-\Delta/2 \le O'< O+\Delta/2} \rho^{(1)}_{\mathbb{L}, \mathbb{S}}(O'), \label{eq:av_overlaps} \\
F^{(2)}_{\mathbb{L},\mathbb{S}}(V;\Delta):= \sum_{V-\Delta/2 \le V'< V+\Delta/2}\rho^{(2)}_{\mathbb{L}, \mathbb{S}}(V'). \label{eq:var_overlaps}
\end{align}

\subsubsection{Fold change of selected gene expression in Fig.\ \ref{fig:logFCs}: }
For the RNA-seq data, we consider the set of cell types $\mathbb{T}$ as
\begin{align}
 \mathbb{T}= \{\rm{HSC}, \rm{CD4}^+ \rm{T}, \rm{CD8}^+ \rm{T}, \rm{NK}, \rm{Mono}, \rm{ATL}\}.
  \end{align}
We compute the fold change in gene $i$, ${\rm FC}_i(t,t_0)$ between type $t,t_0\in\mathbb{T}$ as
\begin{align}\label{eq:F}
  {\rm FC}_{i}(t,t_0):= \dfrac{\overline{R}_i(t)}{\overline{R}_i(t_0)},
\end{align} where $\overline{R}_i(t)$ is the average normalized expression $n_i(s)$ of the RNA-seq data for gene $i$ over all samples with type $t$.
Here, we consider only the genes where $\log_2{\rm FC}_{i}(t,t_0)$ is well-defined;
in other words, we consider the genes $i$ satisfying the following conditions:
(1) a peak of sample type $t$ that intersects the coding region of the gene, and
(2) the same condition for type $t_0$, where the corresponding $k$-th peak satisfies $k\le M$. 
We denote $\mathbb{G}_{t,t_0}(M)$ as the set of all the genes satisfying these conditions.
Then, we focus on the following function quantifying the frequency of the log-fold change
\begin{align} \label{eq:rho}
  \rho_{t,t_0}^M(P):=\dfrac{1}{|\mathbb{G}_{t,t_0}(M)|}\sum_{i\in \mathbb{G}_{t,t_0}(M)}\delta(P,\log_2{\rm FC}_{i}(t,t_0)).
\end{align}
In Fig.\ \ref{fig:logFCs}, we plot the histogram of the frequency of the log-fold change
\begin{align} \label{eq:logFC}
F_{t,t_0}^M(P;\Delta):=\sum_{P-\Delta/2\le P'< P+\Delta/2}\rho_{t,t_0}^M(P').  
\end{align}

\subsubsection{$\Delta \Delta C_t$ 
method in Fig.\ \ref{fig:ATACseq_reads}c:} 
To obtain the threshold cycle for real-time PCR of an mRNA sample (see "Real-time PCR" for details), for sample $s \in \mathbb{S}$, we denote $C_s^{TLL1}$ as the threshold cycle for gene TLL1 and $C_s^{GAPDH}$ for the gene GADPH.
Then, we define the difference $\Delta C_s := C_s^{TLL1} - C^{GAPDH}_s$ and consider the normalized difference as
\begin{align}
    \Delta C_s^0 := \Delta C_s - \min_{s \in \mathbb{S}} \Delta C_s. \label{DDCT}
\end{align}

\subsection*{Data availability}\label{MMsection6}
All ATAC-seq and
RNA-seq data needed to reproduce this study 
will be
deposited at the DNA Data Bank of Japan (DDBJ) under the accession number XXXXX.

\subsection*{Ethics approval and consent to participate}

Experiments using clinical samples were conducted according to the principles expressed in the Declaration of Helsinki and approved by the Institutional Review Board of Kyoto University (permit numbers G310 and G204). ATL patients provided written informed consent for the collection of samples and subsequent analysis.

\subsection*{Author's contributions}
A.~Tanaka: Conceptualization, NGS sample preparation, NGS data analysis, investigation, performing experiments, project administration, generating figures and tables, funding acquisition, and writing original draft. J.I.~Yasunaga: Collecting clinical samples, data investigation, funding acquisition and experimental advices. H.~Ohta and Y.~Ishitsuka: Data investigation, methodology, generating figures, and writing original draft. C.~Onishi and H.~Tanaka: Assisting plasmid preparation, experiments and analyses. N.~Takenouchi, M.~Nakagawa and K.~Koh: Collecting clinical samples. A.~Fujimoto: Assisting NGS data analysis. M.~Matsuoka: Collecting clinical samples, supervision, funding acquisition, project administration, and writing original draft. All authors participated in discussions and interpretation of the data and results.

\begin{acknowledgments}
We thank P.~Karagiannis for proofreading the manuscript and many valuable comments. 
This research was supported by JSPS KAKENHI Grant Numbers JP19K16740 (AT), JP18J40119 (AT), XXXX (MM), and XXXX (JiY) and by a grant from the Naito Foundation (AT). 
\end{acknowledgments}

\bibliography{ATL_211229}

\bibliographystyle{plos2015}

\end{document}